\documentclass[english,aps,manuscript,prb,twocolumn,showpacs,eqsecnum,amsmath,amssymb,reprint,hyperref]{revtex4-1}
\usepackage[T1]{fontenc}
\usepackage[latin9]{inputenc}
\usepackage{babel}
\usepackage{amsmath}
\usepackage{amssymb}
\usepackage{graphicx}
\usepackage[unicode=true,
 bookmarks=false,
 breaklinks=false,pdfborder={0 0 1},backref=false,colorlinks=false]
 {hyperref}
\hypersetup{
 hypertex}
\usepackage{breakurl}

\makeatletter
 
 \@ifundefined{textcolor}{}
 {%
   \definecolor{BLACK}{gray}{0}
   \definecolor{WHITE}{gray}{1}
   \definecolor{RED}{rgb}{1,0,0}
   \definecolor{GREEN}{rgb}{0,1,0}
   \definecolor{BLUE}{rgb}{0,0,1}
   \definecolor{CYAN}{cmyk}{1,0,0,0}
   \definecolor{MAGENTA}{cmyk}{0,1,0,0}
   \definecolor{YELLOW}{cmyk}{0,0,1,0}
 }

\usepackage{bm}
\usepackage{amsfonts}
\hypersetup{backref, 
colorlinks=true,
citecolor=blue,
linkcolor=blue,
filecolor=blue,
urlcolor=blue
}

\makeatother

\begin{document}

\title{Ground state properties of the disordered spin-one Bose-Hubbard model:
a stochastic mean-field theory study }

\author{\emph{Jesus Herazo Warnes}}

\affiliation{\emph{Instituto de F\'{\i}sica Gleb Wataghin, Unicamp, Rua S\'{e}rgio
Buarque de Holanda, 777, CEP 13083-859, Campinas, SP, Brazil}}

\author{\emph{Eduardo Miranda}}

\affiliation{\emph{Instituto de F\'{\i}sica Gleb Wataghin, Unicamp, Rua S\'{e}rgio
Buarque de Holanda, 777, CEP 13083-859, Campinas, SP, Brazil}}

\date{\today}
\begin{abstract}
We study the ground state of the disordered Bose-Hubbard model for spin-1 particles by means of the stochastic mean-field theory. This approach enables the determination of the probability distributions of various physical quantities, such as the superfluid order parameter, the average site occupation number, the standard deviation of the occupation per site and the square of the spin operator per site. We show how a stochastic method, previously used in the study of localization, can be flexibly used to solve the relevant equations with great accuracy. We have determined the phase diagram, which exhibits three phases: the polar superfluid, the Mott insulating and the Bose glass. A complete characterization of the physical properties of these phases has been established.
\end{abstract}


\maketitle

\section{Introduction}

\label{sec:intro}

Systems of cold atoms have become an enormously rich playground for
the study of strongly correlated quantum matter. This era was probably
heralded by the observation of the superfluid to Mott insulator phase
transition is systems of $^{87}Rb$ loaded in optical lattices \cite{greiner02}.
Since then, the possibility of a great amount of control over the
parameters of these systems has attracted the attention of both the
atomic and the condensed matter physics communities \cite{lewenstein07}. 

More recently, the ability to introduce quenched disorder into the
system in a controlled manner has provided researchers with yet another
`knob' to be turned in these studies \cite{lewenstein07}. Disorder
can be incorporated in several ways, namely, through the addition
of laser speckle patterns to the optical lattice potential \cite{horak98,boiron99},
through the creation of a quasi-random optical profile by means of
different laser fields with incommensurate frequencies\cite{roth03,damski03,diener01,Fallani2007,Lucioni11},
by means of randomly trapped atomic `impurities' \cite{gavish05,massignan06},
or even random magnetic fields close to a Feshbach resonance which
can modify locally the scattering length between the atoms \cite{gimperlein05}.
This great flexibility holds a great deal of promise in the study
of the interplay between interactions and disorder, a problem of enormous
importance in condensed matter physics \cite{Lee1985a}.

One other attractive feature of cold atomic systems is the fact that
they can often be very efficiently described by the simple effective
models of condensed matter physics, with much better justification
for the approximations made in arriving at these models. Foremost
among these is the Bose-Hubbard model for spin-zero bosons, which
forms the paradigm for theoretical studies \cite{fisher89,sheshadri93,jaksch98}.
Indeed, in the conventional magnetic traps frequently used, the internal
degrees of freedom (spins) of the atoms are frozen and they can be
described as spinless bosons. However, in purely optical traps the
spins are liberated and the condensates formed depend crucially on
the degeneracy of the atomic spinor \cite{ho98,ohmi98,mukerjee06,Ueda12}.
Again, the usual approximation of retaining only low-energy \emph{s}-wave
scattering between atoms justifies the description of these systems
by means of the Bose-Hubbard model generalized for particles with
spin greater than or equal to one \cite{tsuchiya04}.

A fairly good yet simple treatment of the spin-zero Bose-Hubbard model
is afforded by the so-called `site-decoupled' mean-field theory \cite{fisher89,sheshadri93,van01},
which is able, in particular, to identify the phases of the model
and the topology of the phase diagram at zero temperature. The latter
exhibits (i) a superfluid (SF) phase, characterized by a macroscopic
occupation of the lowest single-particle state (the $\boldsymbol{k}=0$
state in the case of equilibrium) \cite{Yang62} or, equivalently,
by the spontaneous breaking of gauge symmetry \cite{LiebSeiringer2007},
and (ii) a series of Mott insulator (MI) lobes, each characterized
by an integer occupation per site and a vanishing compressibility
due to the presence of an interaction-induced gap. This is in good
qualitative agreement with other more accurate methods (see, e. g.,
\onlinecite{Andersetal2011}). 

This mean-field theory was extended to the spin-1 case and used not
only for the analysis of the ground state of the model \cite{tsuchiya04,kimura05}
but also for finite temperatures \cite{pai08}. In the pioneering
work of reference \onlinecite{tsuchiya04}, the ground state phase diagram
was determined for an antiferromagnetic intra-site interaction. It
was found that, although both superfluid (SF) and Mott insulating
(MI) phases are found, like in the spin-zero case, in contrast to
the latter there are two qualitatively different MI lobes: those with
an odd number of bosons combined in a total spin 1 composite per site,
and even-numbered lobes with a total spin singlet per site. Moreover,
the SF phase was found to have a so-called `polar' structure, corresponding
to a spin-zero condensate. The presence of a non-zero spin per site
can lead to a non-trivial magnetic order. Indeed, it has been proposed
that the MI phases can show spin nematic order, a state with broken
spin rotational symmetry but unbroken time reversal symmetry \cite{imambekov2003}. 

Although the clean Bose-Hubbard models have by now been fairly well
studied, the introduction of quenched disorder poses a much more complex
problem. In the ground-breaking work of reference \onlinecite{fisher89},
scaling arguments were used to address the zero-temperature phase
diagram, the associated quantum phase transitions, and other physical
properties of the disordered spin-zero Bose-Hubbard model. Perhaps
the most important conclusion of that work was the prediction of a
new Bose Glass (BG) phase, which is characterized by localized, insulating
behavior in the absence of an excitation gap and hence, with a finite,
non-zero compressibility. Although the unbiased confirmation of the
existence of this phase by numerical methods is highly non-trivial,
there is by fairly good evidence in favor of it (see, e. g., \onlinecite{krauth91,rapsch99,nikolay04}).

A theoretical study of the effects of disorder in the spin-1 Bose-Hubbard
model was carried out in reference \onlinecite{sanpera11}, in which the
ground state phase diagram was determined by using the Gutzwiller
variational method and a mean-field theory based on the arithmetic
average of the SF order parameter. For the case of diagonal disorder
(i. e., randomness in the local orbital energies), it was found that
the odd-numbered MI lobes are rapidly transformed into a BG regions,
whereas the even-numbered ones are much more robust with respect to
randomness, requiring a much larger strength of disorder before they
also turn into a BG.

The `site-decoupled' mean-field theory mentioned above can be readily
generalized to the disordered case, although its full solution requires
numerical work \cite{sheshadri93,sheshadri95}. A great deal of insight
into this approach can be gained, however, through a simplification
proposed in \onlinecite{bissbort09,bissbort10}. It consists of directing
the focus of attack towards the determination of the probability density
distribution function of local SF order parameters, $P\left(\psi\right)$.
After ignoring correlations among order parameters in nearby sites,
the next step is to establish a mean-field self-consistent condition
to be satisfied by $P\left(\psi\right)$. This method was dubbed Stochastic
Mean Field Theory (SMFT) because it has an immediate formulation as
a stochastic equation. The method offers some advantages over alternative
approaches in that it does not suffer from finite-size effects and
crucially, it allows a great deal of analytical control, specially
over the probability distribution of local quantities. It does have
the drawback of predicting a direct MI-SF transition at weak disorder
without an intervening BG phase, which can be ruled out on firm theoretical
grounds \cite{pollet09}. Nevertheless, despite this shortcoming,
it provides a fairly powerful tool for the analysis of these intricate
disordered systems. Indeed, qualitatively the overall phase boundaries obtained
within SMFT for the zero spin case agree well \cite{bissbort10}
 with Quantum Monte Carlo results in finite lattices \cite{gurarie09}.
 
 We should also mention the important analysis of the disordered spin zero boson problem
 afforded by the real space renormalization group appropriate for strong disorder 
 \cite{altmanetal04,altmanetal08,altmanetal10,iyeretal12}, 
 a powerful tool especially in low dimensions.
 It focused on a quantum rotor representation believed to be equivalent to the Bose-Hubbard 
 Hamiltonian in the limit of a large number of bosons per site. The system was thoroughly 
 characterized in one spatial dimension \cite{altmanetal04,altmanetal08,altmanetal10}, 
 which is special since there can be 
 only quasi-long range (power-law) superfluid order in the ground state.
 Like the clean case, the quantum superfluid-insulator transition belongs to the Kosterlitz-Thouless 
 universality class. In the disordered case, however, this transition can occur at arbitrarily
 weak interactions, which sets it apart for its higher-dimensional counterparts. The insulating phase has
 the expected features (i.e. a finite compressibility) of a Bose glass for generic disorder. 
 Other types of disorder with special particle-hole
 symmetry properties were also considered, in which case the insulator can have
 vanishing (the so-called Mott glass) or infinite compressibility (dubbed a random singlet glass).
 More recently, this approach has been extended to two dimensions \cite{iyeretal12}, 
 in which case the transition is governed by a more conventional unstable fixed point at finite interaction strength. 
 However, this was confined to the non-generic particle-hole symmetric disorder 
 that does not give rise to a Bose glass phase. It should be mentioned that all fixed points
 found show finite effective disorder, which renders the method less conclusive than
 at other infinite disorder fixed points for which the method is asymptotically exact.

The aim of this paper is to analyze the disordered spin-1 Bose-Hubbard
model with the tools of the SMFT \cite{bissbort09,bissbort10}. We
will focus our attention on the antiferromagnetic interaction case
only. We have found that the phase diagram of reference \onlinecite{sanpera11}
is well captured by this simplified approach. Furthermore, we are
also able to find a number of distribution functions of local quantities
which offer a great deal of insight into the nature of the various
phases, namely, the local spinor order parameters, the average and
the standard deviation of the site occupation, and the total spin
per site. Finally, by analyzing the behavior of the system both as
a function of interactions and as a function of disorder strength
we track the hierarchical transformation of the MI lobes into BG phases.
It should be mentioned that, unlike in the original application of
the SMFT \cite{bissbort09,bissbort10}, we use an alternative stochastic
approach to solve the SMFT equations which was first suggested in
reference \onlinecite{anderson73} and used extensively in \onlinecite{aguiaretal2003} 
and proves to be quite efficient and flexible.

The paper is organized as follows. In Section \ref{sec:model_h},
we present the model and review the phase diagram both in the clean
and disordered cases. In Section \ref{sec:smft}, we explain how the
SMFT is defined and also describe our strategy for solving the corresponding
equations. We then present our results in Section \ref{sec:results}.
We wrap up with some conclusions in Section \ref{sec:conclusion}.

\section{The model}

\label{sec:model_h}

We will focus on a generalized disordered Hubbard model for bosons
with total spin $F=1$ \cite{tsuchiya04}
\begin{eqnarray}
H & = & -t\sum_{\langle i,j\rangle}\sum_{\alpha}a_{i\alpha}^{\dagger}a_{j\alpha}^{\phantom{\dagger}}\nonumber \\
 & + & \sum_{i}\sum_{\alpha}(\epsilon_{i}-\mu)a_{i\alpha}^{\dagger}a_{i\alpha}^{\phantom{\dagger}}\nonumber \\
 & + & \dfrac{U_{0}}{2}\sum_{i}\sum_{\alpha,\beta}a_{i\alpha}^{\dagger}a_{i\beta}^{\dagger}a_{i\beta}^{\phantom{\dagger}}a_{i\alpha}^{\phantom{\dagger}}\nonumber \\
 & + & \dfrac{U_{2}}{2}\sum_{i}\sum_{\alpha,\beta,\gamma,\delta}a_{i\alpha}^{\dagger}a_{i\gamma}^{\dagger}\boldsymbol{S}_{\alpha\beta}\cdot\boldsymbol{S}_{\gamma\delta}a_{i\delta}^{\phantom{\dagger}}a_{i\beta}^{\phantom{\dagger}},\label{eq:hamiltonian}
\end{eqnarray}
where $a_{i\alpha}^{\dag}$ creates a bosonic atom with spin projection
$\alpha\in\left\{ -1,0,1\right\} $ in an optical lattice Wannier
function centered on the site $i$, $t$ is a nearest-neighbor hopping
amplitude, $\mu$ is the chemical potential (we will work in the grand-canonical
ensemble), $U_{0}$ and $U_{2}$ are local (intra-site) coupling constants
for spin-independent and spin-dependent interactions, respectively,
and $\boldsymbol{S}=S_{x}\boldsymbol{\hat{x}}+S_{y}\boldsymbol{\hat{y}}+S_{z}\boldsymbol{\hat{z}}$
are the spin-1 matrices given by

\begin{eqnarray}
S_{x} & = & \frac{1}{\sqrt{2}}\begin{pmatrix}0 & 1 & 0\\
1 & 0 & 1\\
0 & 1 & 0
\end{pmatrix},\ S_{y}=\dfrac{i}{\sqrt{2}}\begin{pmatrix}0 & -1 & 0\\
1 & 0 & -1\\
0 & 1 & 0
\end{pmatrix},\nonumber \\
S_{z} & = & \begin{pmatrix}1 & 0 & 0\\
0 & 0 & 0\\
0 & 0 & -1
\end{pmatrix}.\label{eq:spinoperators}
\end{eqnarray}
The interaction coupling constants can be related to the $s$-wave
scattering lengths of two bosons in vacuum with total spin 0 ($a_{0}$)
and 2 ($a_{2}$) (the symmetric nature of their wave function forbidding
$s$-wave processes with total spin 1) \cite{tsuchiya04}
\begin{eqnarray}
U_{0} & = & \frac{4\pi\hbar^{2}}{3M}\left(a_{0}+2a_{2}\right)I_{4},\label{eq:U0}\\
U_{2} & = & \frac{4\pi\hbar^{2}}{3M}\left(a_{2}-a_{0}\right)I_{4},\label{eq:U2}
\end{eqnarray}
where $M$ is the boson mass and $I_{4}$ is the integral of the fourth
power of the Wannier wave function. As per the usual nomenclature,
the spin-dependent interaction is called ferromagnetic, when $U_{2}<0$
(i.e., $a_{2}<a_{0}$) and antiferromagnetic if $U_{2}>0$ (i.e.,
$a_{2}>a_{0}$) \cite{ho98}. On-site disorder is introduced through
the parameters $\epsilon_{i}$, which are taken to be random quantities
with no spatial correlations. Although several models of disorder
may be considered, for simplicity we chose $\epsilon_{i}$ to be distributed
according to a uniform distribution of width $2\Delta$.

It is useful to introduce the single-site operators for the total
number of bosons $n_{i}$ and total spin $\boldsymbol{S}_{i}$, 
\begin{eqnarray}
\hat{n}_{i} & = & \sum_{\alpha}a_{i\alpha}^{\dagger}a_{i\alpha}^{\phantom{\dagger}},\label{eq:numberop}\\
\boldsymbol{S}_{i} & = & \sum_{\alpha,\beta}a_{i\alpha}^{\dagger}\boldsymbol{S}_{\alpha\beta}a_{i\beta}^{\phantom{\dagger}},\label{eq:spinop}
\end{eqnarray}
in terms of which the Hamiltonian \eqref{eq:hamiltonian} can be rewritten
as
\begin{eqnarray}
H & = & -t\sum_{\alpha}\sum_{\langle i,j\rangle}a_{i\alpha}^{\dagger}a_{j\alpha}^{\phantom{\dagger}}+\dfrac{U_{0}}{2}\sum_{i}\hat{n}_{i}(\hat{n}_{i}-1)\nonumber \\
 & + & \dfrac{U_{2}}{2}\sum_{i}(\boldsymbol{S}_{i}^{2}-2\hat{n}_{i})+\sum_{i}(\epsilon_{i}-\mu)\hat{n}_{i}.\label{eq:hamiltonianb}
\end{eqnarray}

In the clean limit, the model exhibits two phases: a superfluid phase
(SF) and several lobes of Mott insulating (MI) behavior \cite{tsuchiya04}.
The SF is the so-called polar state, characterized by a spin-zero
Bose-Einstein condensate (BEC). As any superfluid state, it can be
characterized by the appearance of a non-zero value of $\left\langle a_{i\alpha}\right\rangle $
for some component $\alpha$, and this situation corresponds to the
spontaneous breaking of gauge symmetry \cite{LiebSeiringer2007}.
A convenient, albeit not unique choice for the order parameter structure
of the polar state is $\psi_{-1}=\psi_{1},\psi_{0}=0$. The MI lobes,
on the other hand, can be classified in two categories: those that
correspond to an odd number $n=\sum_{i}\left\langle \hat{n}_{i}\right\rangle /N$
of bosons per site (here, $N$ is the number of sites), which combine
to form a spin-1 object on each site, and lobes in which each site
has a 0-spin even $n$ combination. Generically, the even-numbered
lobes are more stable and tend to occupy a larger fraction of the
$\mu$ versus $t$ phase diagram as compared to the nearby odd-numbered
lobes, which are smaller and disappear altogether for $U_{2}/U_{0}\ge0.5$.
In addition, for $U_{2}/U_{0}<r_{c}\approx0.2$ the even-numbered
MI-SF quantum phase transition is first order in character, as opposed
to the odd-numbered one which is always continuous. For $U_{2}/U_{0}\ge r_{c}$,
all MI-SF transitions are continuous \cite{kimura05,pai08,sanpera11}.
Finally, the MI is characterized by a vanishing compressibility $\kappa=\partial n/\partial\mu$,
which in contrast remains non-zero in the SF.

Once disorder is turned on, a new phase appears: the Bose glass (BG)
\cite{fisher89,sanpera11}. The latter is not a SF and therefore the
order parameter is zero for any value of $\alpha$. More precisely,
since $\psi_{i\alpha}$ becomes a random quantity in the disordered
system, its distribution is given by $P_{\alpha}\left(\psi_{\alpha}\right)=\delta\left(\psi_{\alpha}\right)$
for any $\alpha$. However, unlike the MI phase, the charge excitation
spectrum is gapless and the fluid is compressible: $\partial n/\partial\mu\neq0$.
A full specification of all phases thus requires the computation of
the order parameter distribution and the compressibility. It should
be noted that there was a long-standing controversy over whether the
topology of the phase diagram is such as to allow a direct transition
from a MI to a SF, without passing through an intervening BG phase.
Scaling arguments suggested that such a direct MI-SF transition is
unlikely in the presence of disorder \cite{fisher89}. However, numerical
results proved to be inconclusive (see, e. g., \onlinecite{krauth91} and
\onlinecite{nikolay04}). More recently, extreme-statistics arguments have
been used to show that there are necessarily extended Lifshitz regions
of gapless particle-hole excitations at the SF phase boundary \cite{pollet09}.
Therefore, it seems clear now that there is always a BG phase adjacent
to the SF and a direct MI-SF is not possible in the disordered case.

\section{The stochastic mean field theory}

\label{sec:smft}

The superfluid-Mott insulator transition of lattice bosons can be
qualitatively captured by a standard mean field approach which is
based upon decoupling the hopping term of the Hamiltonian (Eq. (\eqref{eq:hamiltonianb}))
as \cite{fisher89,van01,tsuchiya04} 
\begin{equation}
a_{i\alpha}^{\dagger}a_{j\alpha}^{\phantom{\dagger}}\simeq\psi_{i\alpha}^{*}a_{j\alpha}^{\phantom{\dagger}}+a_{i\alpha}^{\dagger}\psi_{j\alpha}-\psi_{i\alpha}^{*}\psi_{j\alpha},\label{eq:mfdecoupling}
\end{equation}
where $\psi_{i\alpha}=\langle a_{i\alpha}\rangle$ and we are neglecting
second-order fluctuations ${\cal O}\left(a_{j\alpha}^{\phantom{\dagger}}-\langle a_{j\alpha}\rangle\right)^{2}$.
The order parameters $\psi_{i\alpha}$ have to be determined self-consistently.
This is achieved by focusing on the decoupled single-site Hamiltonians
generated after \eqref{eq:mfdecoupling} is applied to \eqref{eq:hamiltonianb}

\begin{eqnarray}
h_{i} & = & -\sum_{\alpha}\left(\eta_{i\alpha}a_{i\alpha}^{\dag}+\eta_{i\alpha}^{*}a_{i\alpha}-\psi_{i\alpha}\eta_{i\alpha}^{*}\right)+(\epsilon_{i}-\mu)n_{i}\nonumber \\
 & + & \frac{U_{0}}{2}n_{i}(n_{i}-1)+\frac{U_{2}}{2}(\textbf{S}_{i}^{2}-2n_{i}),\label{eq:singlesiteham}
\end{eqnarray}
where 
\begin{equation}
\eta_{i\alpha}=t\sum_{j=1}^{Z}\psi_{j\alpha},\label{eq:etadefinition}
\end{equation}
where $Z$ is the lattice coordination number. Once $\langle a_{i\alpha}\rangle\left(\epsilon_{i},\left\{ \psi_{j\beta}\right\} \right)$
is determined from \eqref{eq:singlesiteham} (note the dependence
on the local site energy $\epsilon_{i}$ and adjacent order parameters
$\psi_{j\alpha}$), self-consistency is assured if we impose that
\begin{equation}
\psi_{i\alpha}=\langle a_{i\alpha}\rangle\left(\epsilon_{i},\left\{ \psi_{j\beta}\right\} \right).\label{eq:self-consistency}
\end{equation}

Complete lattice self-consistency requires solving the large set of
coupled equations defined by Eqs. \eqref{eq:self-consistency}. A
considerable simplification can be achieved if we neglect \emph{spatial
correlations} between sites. This defines the so-called \emph{stochastic
mean field theory} (SMFT) of references \onlinecite{bissbort09,bissbort10},
which was originally proposed for the spin-0 model, but which we will
now describe for the generic spinful case. For this, we note that
$\langle a_{i\alpha}\rangle$ depends on the other order parameters
only through $\eta_{i\alpha}$, see Eq. \eqref{eq:singlesiteham}.
We thus look for the distributions of local order parameters $P_{\alpha}\left(\psi_{\alpha}\right)$
by first finding the distributions of $\eta_{i\alpha}$, $Q_{\alpha}\left(\eta_{\alpha}\right)$,
which are induced by $P_{\alpha}\left(\psi_{\alpha}\right)$ through
Eq. \eqref{eq:etadefinition}, \emph{neglecting spatial correlations
between different nearest neighbors}. Next, we use the fixed function
$\langle a_{i\alpha}\rangle\left(\epsilon_{i},\eta_{i\alpha}\right)$,
which usually has to be obtained numerically, to generate the induced
distributions $A_{\alpha}\left(\langle a_{i\alpha}\rangle\right)$.
Finally, self-consistency is obtained by imposing that $A_{\alpha}\left(x\right)=P_{\alpha}\left(x\right)$.
Despite its approximate nature, this approach has been shown to be
able to capture all the phases of the spin-0 model \cite{bissbort09,bissbort10}.

The procedure described above for the SMFT can be implemented as a
non-linear integral equation for the sought distributions $P_{\alpha}\left(\psi_{\alpha}\right)$,
which can then be solved numerically on a discrete mesh. This was
the approach used in references \onlinecite{bissbort09,bissbort10}. We
opted instead to use an importance sampling method, akin to the Monte
Carlo method, as originally proposed for the self-consistent theory
of localization \cite{anderson73,aguiaretal2003}. The method can be described as
follows. We start from a sample of random values for $\psi_{i\alpha}^{\left(0\right)}\ \left(i=1,...,N_{s}\right)$
which are drawn from an initial guess for the sought distributions,
$P_{\alpha}^{\left(0\right)}\left(\psi_{\alpha}\right)$. The method
is very robust with respect to the choice of this initial guess, so
we can start with a uniform distribution. From this initial sample,
we generate a corresponding initial sample for $\eta_{i\alpha}\ \left(i=1,...,N_{s}/Z\right)$
using Eq. \eqref{eq:etadefinition}, which may be viewed as an initial
guess $Q_{\alpha}^{\left(0\right)}\left(\eta_{\alpha}\right)$. Using
the latter and a corresponding sample of $N_{s}/Z$ values of $\epsilon_{i}$
drawn from its (given) distribution, we can then find (numerically
solving for $\langle a_{i\alpha}\rangle$ according to \eqref{eq:singlesiteham})
a new sample of values of the order parameter $\psi_{i\alpha}^{\left(1\right)}\ \left(i=1,...,N_{s}/Z\right)$,
which can be viewed as drawn from an improved distribution $P_{\alpha}^{\left(1\right)}\left(\psi_{\alpha}\right)$.
The sample size will have decreased by a factor of $1/Z$ from the
previous iteration. For further iterations, we can enlarge this smaller
sample to the original size by replicating it $Z$ times and reshuffling
it. For a large enough value of $N_{s}$ this leads to negligible
errors, which can be checked by studying the dependence of the final
results on $N_{s}$. The procedure can then be iterated many times
until sample-to-sample variations are negligible, which can be verified,
for example, by computing sample features such as its mean and variance.
A numerical estimate of the converged distribution is then obtained
from a histogram of the last several iterations. Furthermore, histograms
of any local quantities can also be easily generated.

\begin{figure}
\includegraphics[scale=0.25]{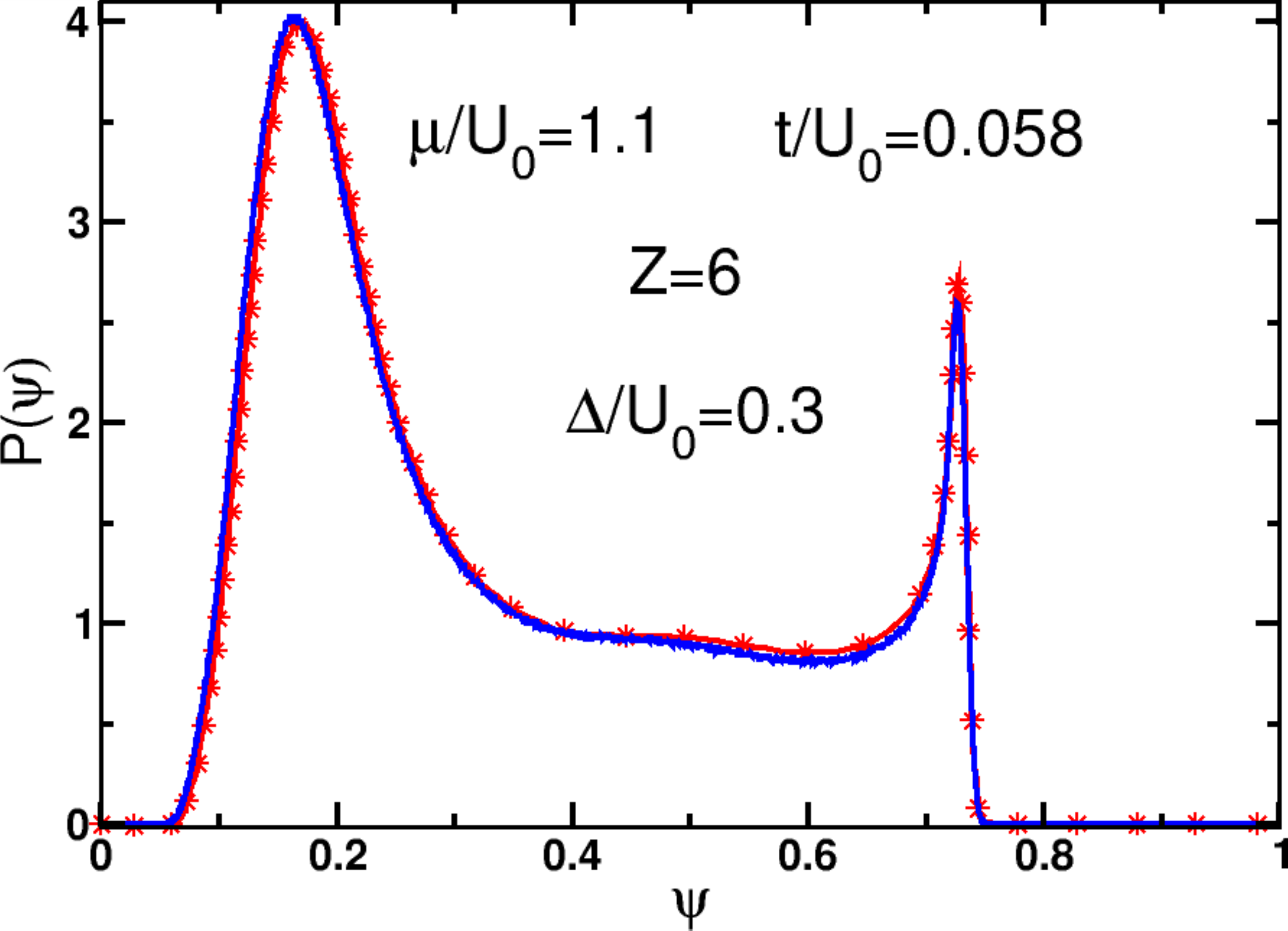} \caption{\label{fig:spin0} (Color online) Order parameter distribution for
the spin zero disordered Hubbard model inside the Bose-Einstein condensed
phase obtained within SMFT for $\mu/U_{0}=1.1$, $t/U_{0}=0.058$,
$\Delta/U_{0}=0.3$, and $Z=6$. The full blue line was obtained with
the importance sampling method used in this paper and the symbols
come from solving directly the integral equation (extracted from reference
\onlinecite{BissbortThesis}).}
\end{figure}

We have compared the two different methods of solving the SMFT equations, namely, 
the stochastic method proposed in this paper and the direct solution
of the integral equation used in references \onlinecite{bissbort09,bissbort10}, 
in the spin zero case. In Fig.~\ref{fig:spin0},
we show a typical result for the order parameter distribution $P\left(\psi\right)$
obtained by these two methods, within the disordered Bose-Einstein
condensed phase of the spin zero disordered Hubbard model. The importance
sampling approach used $N_{s}=360,000$ and it took 30 iterations
for the convergence to be achieved, after which 70 more iterations
were obtained to generate the final statistics. The agreement is remarkable
and makes us confident that the method is reliable. In fact, we achieved
enough accuracy with $N_{s}=60,000$, 40 iterations before convergence
and 60 more to gather enough statistics.

Finally, we should mention that the direct computation of the compressibility
within SMFT points to a direct transition between the MI and the SF
at weak disorder \cite{bissbort09,bissbort10}, which is at odds with
the rigorous results of \onlinecite{pollet09}. The SMFT is thus incapable
of capturing the rare regions of gapless charge excitations close
to the SF phase that preclude such a direct transition. Arguments
have been given in \onlinecite{bissbort10}, showing how to reinterpret
the SMFT phase diagram in order to correct for this failure. Nevertheless,
it should be kept in mind that the direct calculation of $\kappa$
does not show the expected behavior.

In the next Section, we will show the results of the SMFT as applied
to the disordered spin-1 model of Eq.~\eqref{eq:hamiltonian}.

\section{Results}

\label{sec:results}

We now present the results of applying the SMFT to the spin-1 model
of Eq.~\eqref{eq:hamiltonian} at $T=0$. In all the following results,
we have fixed the spin-dependent interaction coupling to be antiferromagnetic
with $U_{2}/U_{0}=0.3$ and $Z=4$. In Sections~\ref{sec:order_parameter}
to \ref{sec:spin} we fix the disorder strength at $\Delta/U_{0}=0.3$.
In Section~\ref{sec:disorder_function}, we study the behavior of
the system at fixed $\mu$ and $t$ as a function of the disorder.

\subsection{Order parameter}

\label{sec:order_parameter}

\begin{figure}
\includegraphics[scale=0.3]{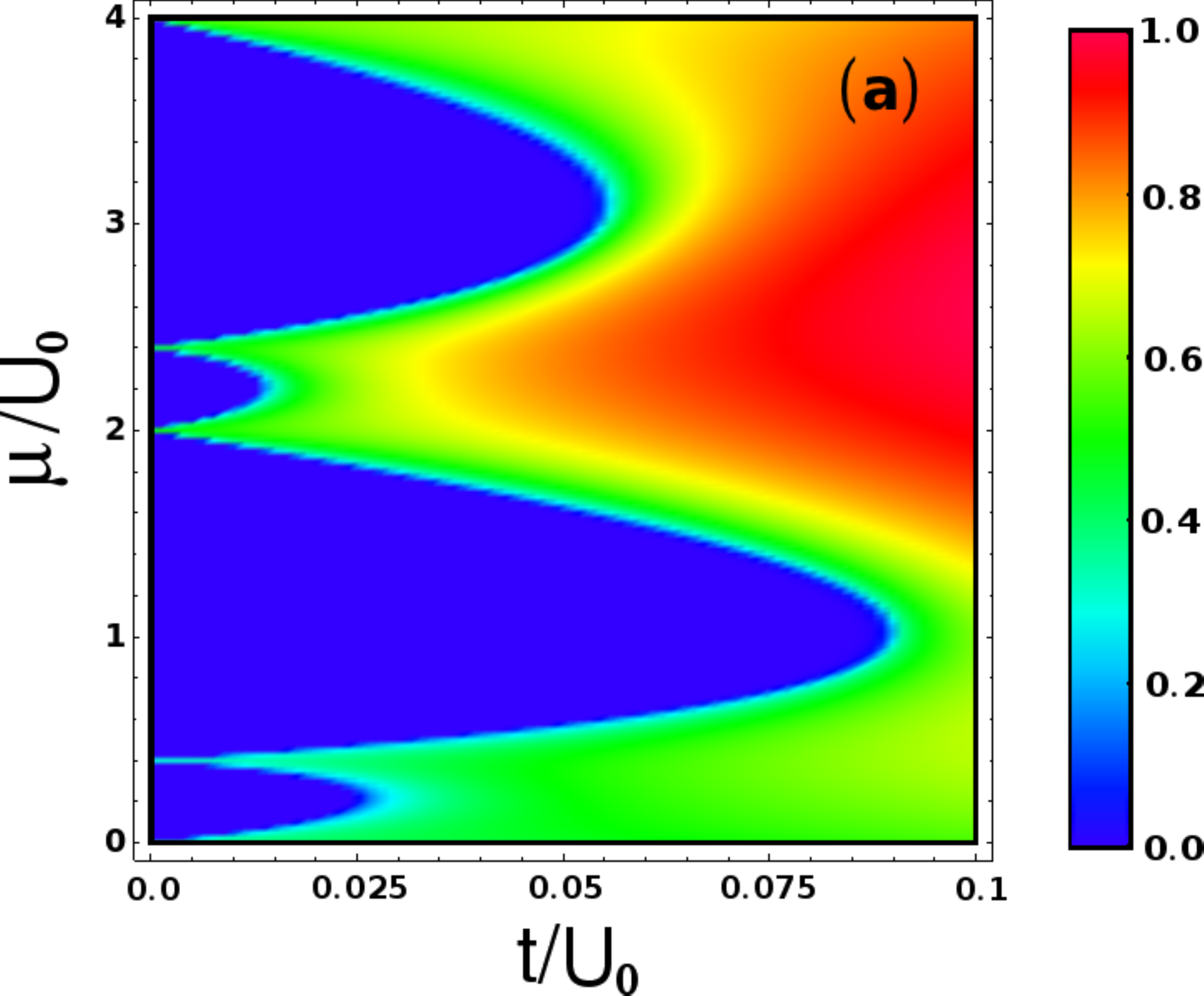} 

\includegraphics[scale=0.3]{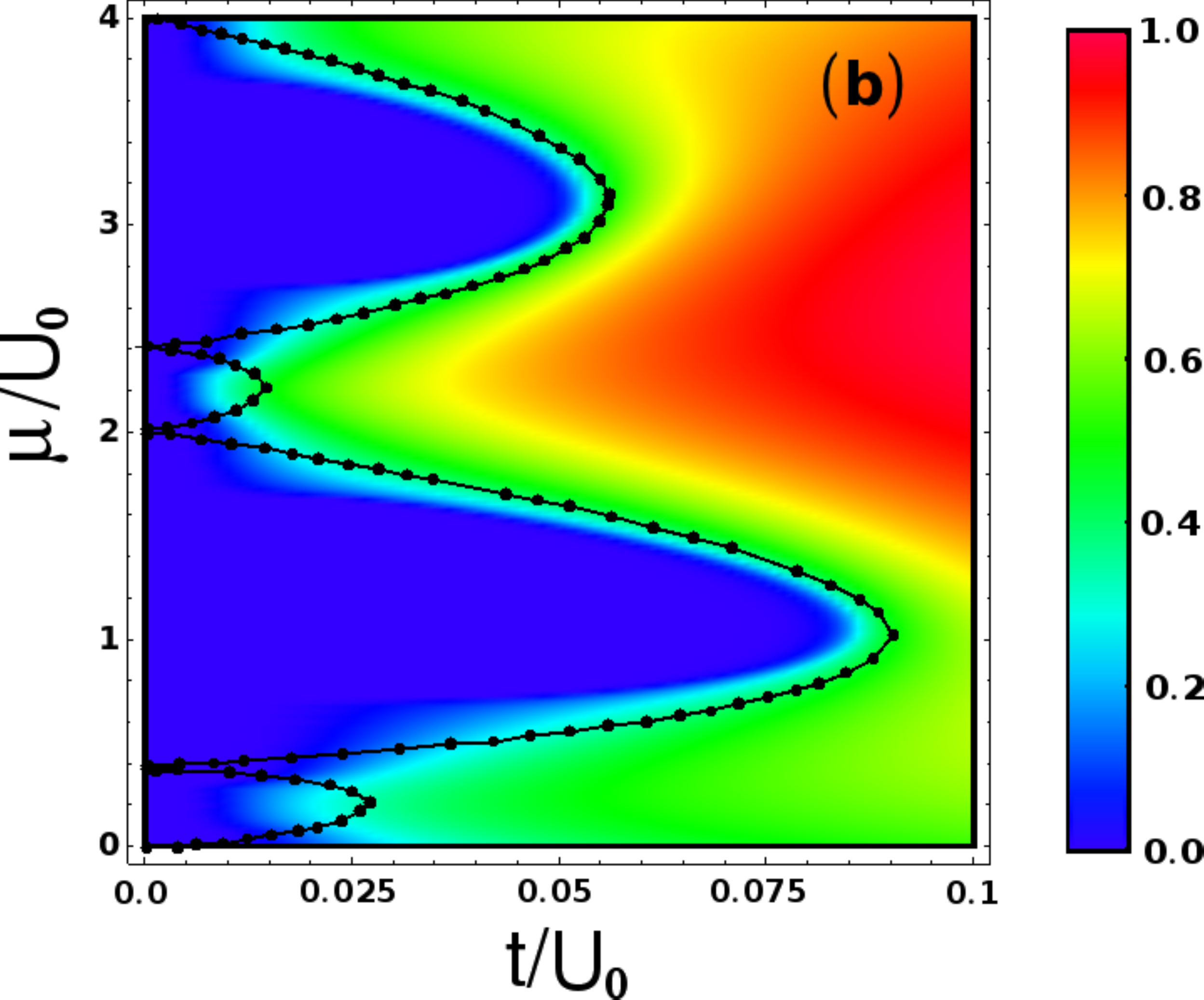} \caption{\label{fig:phi1phasediag}(Color online) Average value of the $\psi_{1}$
component of the order parameter in the $(t,\mu)$ plane for the (a)
clean and (b) disordered ($\Delta/U_{0}=0.3$) cases. In (b) the boundaries of the clean Mott lobes of (a) are shown as black dotted lines.}
\end{figure}

We focus first on the behavior of the order parameter. The clean polar
SF phase is characterized by an order parameter structure in which,
for a particularly convenient gauge choice, $\psi_{-1}=\psi_{1}$
and $\psi_{0}=0$ \cite{tsuchiya04}. This phase corresponds to a
spin-zero condensate, as can be easily checked. Fig.~\ref{fig:phi1phasediag}(a)
shows the value of $\psi_{1}$ as a color scale plot in the $\mu$
vs $t$ phase diagram. The Mott lobes can be clearly identified and
also the fact that the even-numbered ones occupy a much larger portion
of the phase diagram. The transition from SF to both types of MI is continuous
for this value of $U_2/U_0$ \cite{sanpera11}.

We then add disorder ($\Delta/U_{0}=0.3$) within a SMFT treatment.
We were able to find only converged solutions with $P_{1}\left(x\right)=P_{-1}\left(x\right)$
and $P_{0}\left(\psi_{0}\right)=\delta\left(\psi_{0}\right)$. In
other words, although the order parameter is now a random quantity,
it still preserves the same component structure as in the clean case.
We have thus produced a color scale plot of the average value of $\psi_{1}$
in the same $\mu$ vs $t$ phase diagram, as can be seen in Fig.~\ref{fig:phi1phasediag}(b).
This can be viewed as an order parameter for the SF phase in the disordered
system. The transitions remain continuous within our accuracy. The boundaries
of the Mott lobes of the clean case are shown as black dotted lines for comparison.
There is a clear suppression of the regions with a vanishing order parameter (blue regions),
except at the wedges that separate the clean Mott lobes, where superfluid order is suppressed.
A definite characterization of the non-superfluid regions will be carried out later,
when we show the results for the compressibility in Section \ref{sec:compressibility}.
We anticipate that the large even-numbered lobes will retain their Mott insulating character.
There is a clear suppression of these lobes by disorder, which are
seen to become narrower and to extend up to smaller maximum values
of the hopping amplitude as compared to the clean case. In contrast, the odd-numbered
lobes will be shown to have transformed into the BG phase with a finite compressibility.
Their shape is completely deformed by disorder. The conclusion is that
the even-numbered MI lobes are more resilient to the effects of disorder.
Just like in the clean case, the positive value
of $U_{2}$, which stabilizes the even occupation, also acts to localize
the bosons more strongly, thus protecting the MI phase against weak
disorder. Finally, the small hopping SF wedges that exist between
the MI lobes in the clean system are also suppressed by disorder and
go into the BG phase.

\begin{figure}
\includegraphics[scale=0.25]{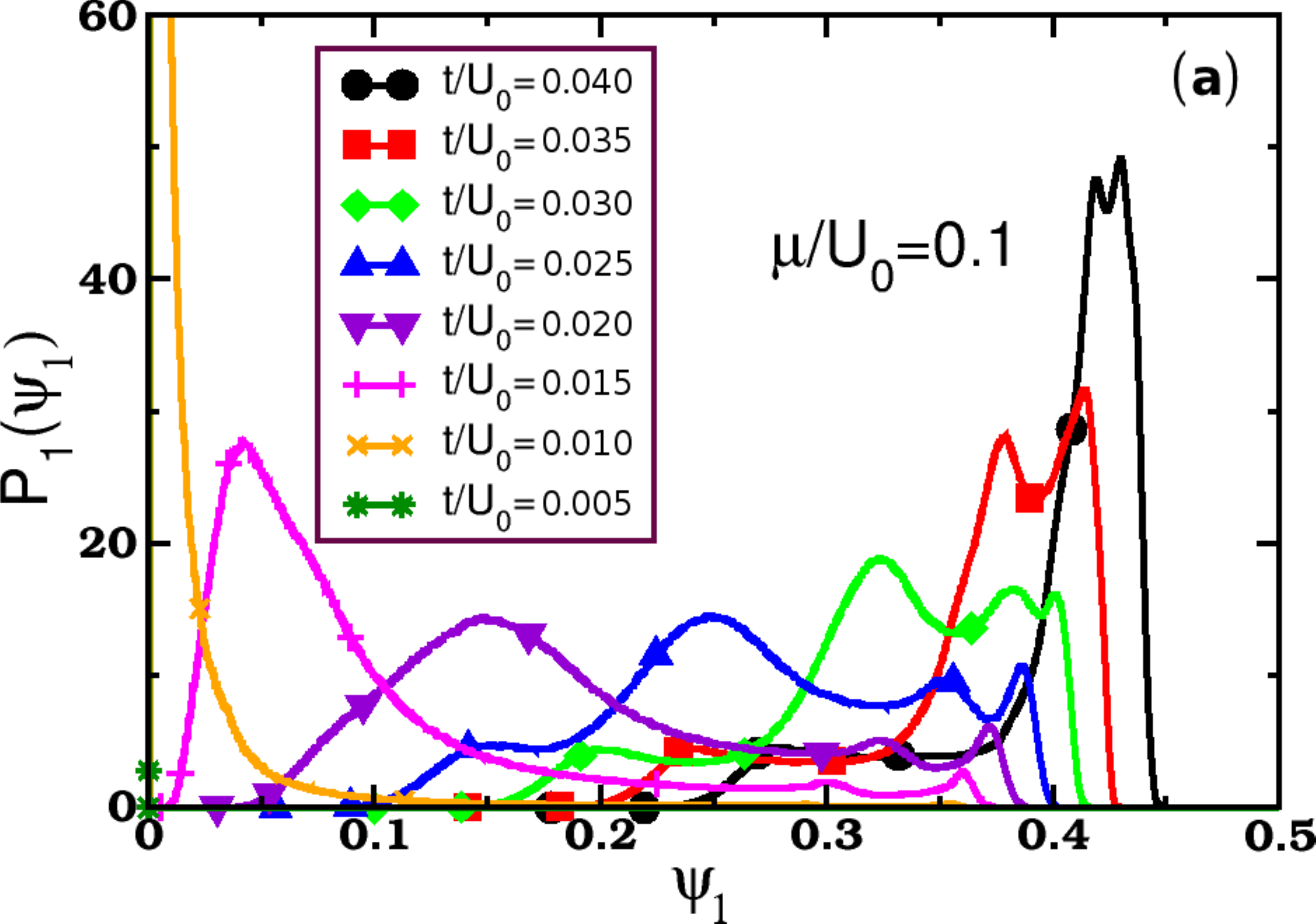}

\includegraphics[scale=0.25]{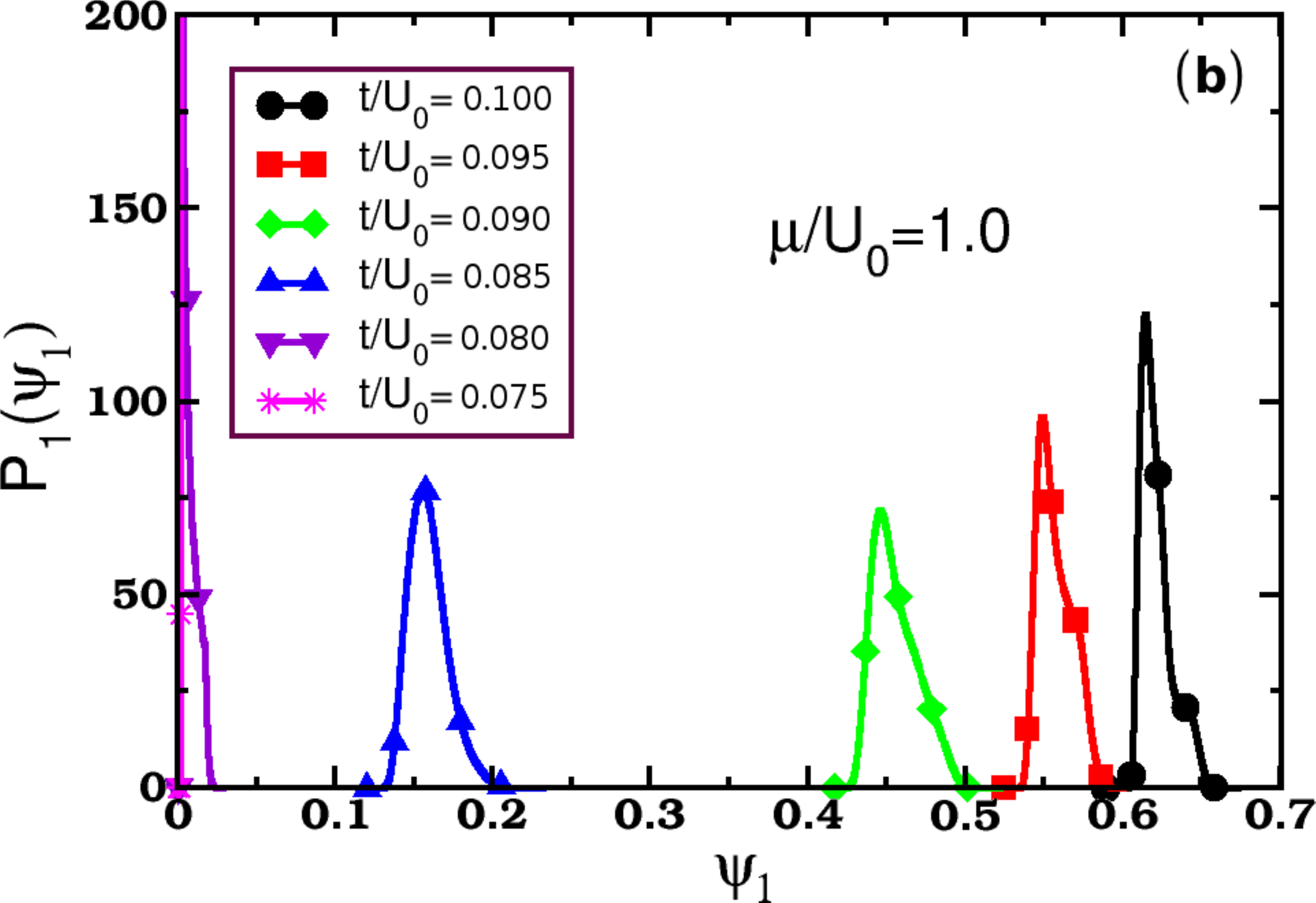} \caption{\label{fig:phi1dist} (Color online) Probability distributions for
the components of the order parameter $P_{1}\left(\psi_{1}\right)$
for several values of the hopping amplitude and two values of the
chemical potential: (a) $\mu/U_{0}=0.1$ and (b) $\mu/U_{0}=0.1$.
The disorder is set to $\Delta/U_{0}=0.3$. As $t$ decreases, the
system goes from a disordered polar SF to a BG in (a) and to a MI
in (b).}
\end{figure}

The full distribution functions $P_{1}\left(\psi_{1}\right)$ are
shown in Fig.~\ref{fig:phi1dist} for two different values of the
chemical potential and several values of the hopping amplitude (again
$\Delta/U_{0}=0.3$). In Fig.~\ref{fig:phi1dist}(a), the value of
the chemical potential is $\mu/U_{0}=0.1$, which corresponds to the
$n=1$ MI lobe in the clean case at small $t$. As $t$ is decreased
the disordered system goes from a polar SF to a BG phase. It is interesting
to note that the distribution is fairly narrow deep in the SF and
become increasingly broader and distorted as $t$ decreases, while
at the same time its weight shifts towards small values of $\psi_{1}$.
In particular, for values of $t$ close to the BG (see, e. g., $t/U_{0}=0.015$
and $0.01$), the distribution shows a very skewed shape with a peak
at an increasingly smaller $\psi_{1}$ and a long tail for larger values
of the order parameter. Eventually, it tends towards a delta function
at $\psi_{1}=0$ inside the BG phase, barely visible on the scale
of the figure at $t/U_{0}=0.005$. This generic behavior is also observed
in the SF to BG transition of the spin-zero model \cite{bissbort09}.

In Fig.~\ref{fig:phi1dist}(b), the chemical potential is set to
$\mu/U_{0}=1.0$, which in the clean system gives rise to the $n=2$
MI lobe at small $t$. As we add disorder, the system can be tuned
from the SF to a disordered MI phase. The order parameter distribution
shows a markedly different behavior when compared to the $\mu/U_{0}=0.1$
case. Indeed, it retains a fairly narrow shape as $t$ is decreased,
while shifting its weight to ever smaller values of $\psi_{1}$, eventually
tending to a delta function at zero within the MI lobe.

\subsection{Compressibility}

\label{sec:compressibility}

As was mentioned before, it is essential to analyze the behavior of
the compressibility $\kappa=\dfrac{\partial n}{\partial\mu}$ in order
to obtain a complete characterization of the phases: this quantity
is finite in both the SF and the BG phases but vanishes in the MI
\cite{fisher89}.

\begin{figure}
\includegraphics[scale=0.3]{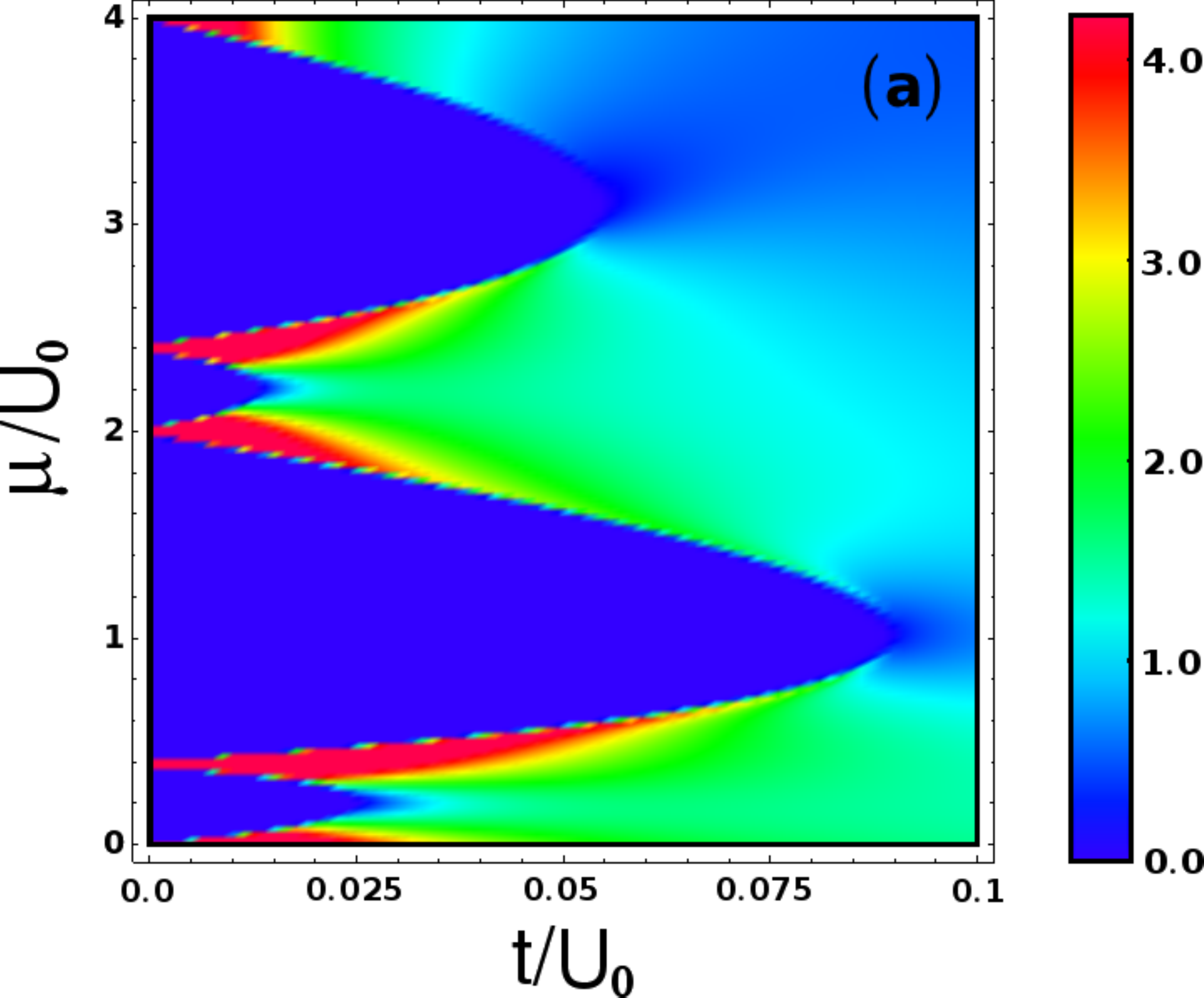}

\includegraphics[scale=0.3]{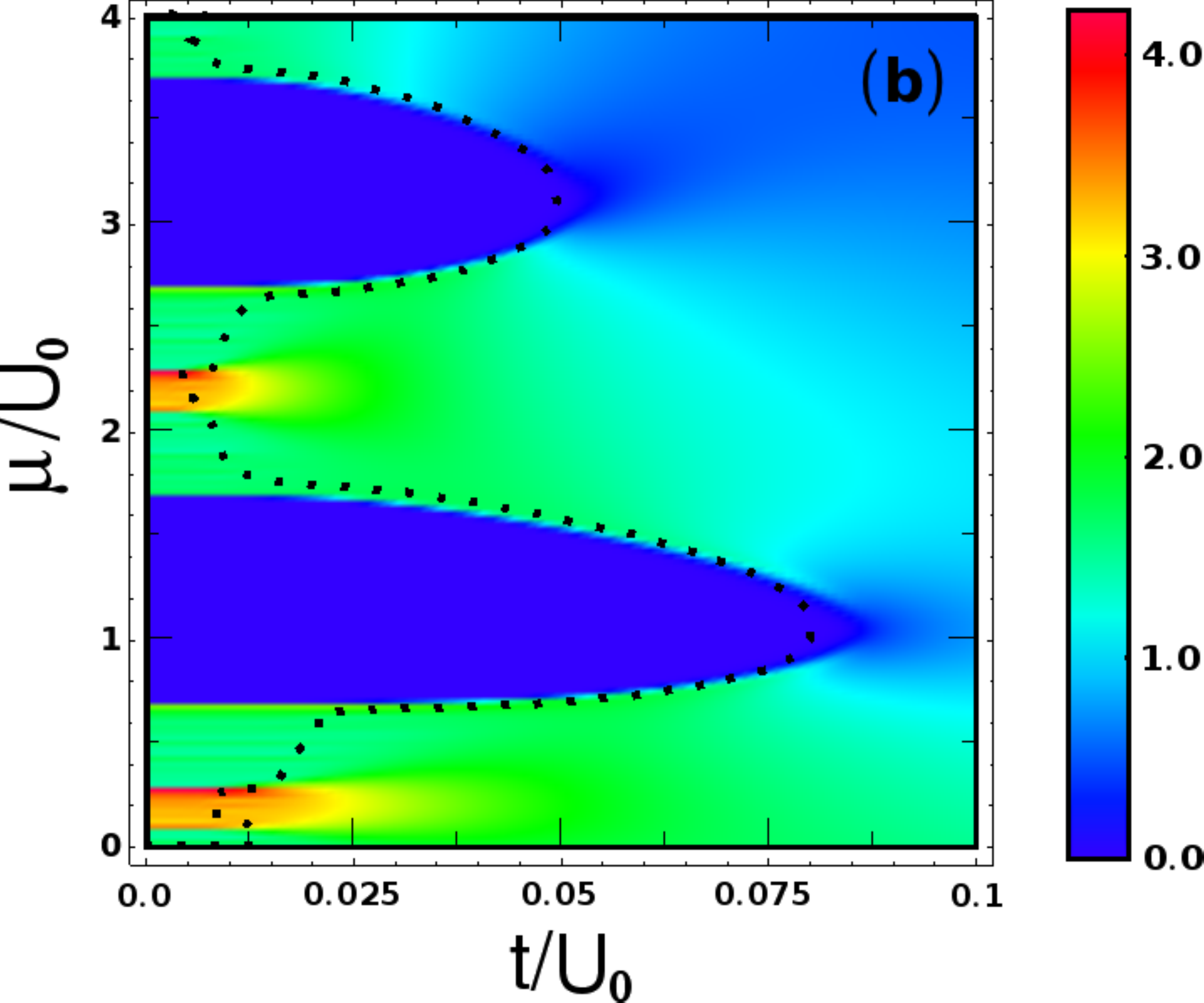} \caption{\label{fig:kappaphasediag}(Color online) Compressibility in the $(t,\mu)$
phase diagram. (a) Clean case and (b) disordered case ($\Delta/U_{0}=0.3$). The dotted line delineates the regions characterized by a vanishing order parameter (see Fig.~\ref{fig:phi1phasediag}(b) }
\end{figure}

Fig.~\ref{fig:kappaphasediag} is shows $\kappa$ in the $\mu$ vs
$t$ plane using a color scale for both the clean and disordered cases
($\Delta/U_{0}=0.3$). In the clean case (Fig.~\ref{fig:kappaphasediag}(a)),
the compressibility is zero inside the MI lobes and non-zero in the
SF phase. Note that, as $t\to0$, the SF phase disappears and the
MI lobes are characterized by integer site occupancies, the latter
then giving rise to a series of steps of increasing value as $\mu$
increases. As a result, the compressibility diverges as one crosses
from one lobe to the next at the $t=0$ line, since there is a jump
in $n$. Therefore, large values of $\kappa$ cluster around these
transitions in the small $t$ region (red color in the figure). In
that figure, we have arbitrarily set $\kappa=4.22$ to compressibilities
equal to or greater than this value.

The compressibility of the disordered system is shown in Fig.~\ref{fig:kappaphasediag}(b).
The regions with zero order parameter from Fig.~\ref{fig:phi1phasediag}(b) have been delineated 
as the dotted lines.
As can be seen, the compressibility remains zero in large portions of the phase diagram. 
These regions thus have both vanishing compressibility and order parameter and 
correspond to even-numbered MI lobes,
cf. Fig.~\ref{fig:phi1phasediag}. Thus, as in the case of the
spin-zero model \cite{bissbort09,bissbort10}, the SMFT predicts a
direct MI-SF transition at this value of disorder, which is an artifact
of the approximation used \cite{pollet09}.

In contrast, however, the small regions which were the odd-numbered MI lobes 
in the clean case now exhibit a non-zero
$\kappa$ once disorder is added. In other words, these clean MIs
are completely destroyed by this amount of randomness and become BGs.
It should be said that even within the SF phase the compressibility
can become very small (blueish regions), even though it remains non-zero
everywhere in the SF.

\begin{figure}
\includegraphics[scale=0.25]{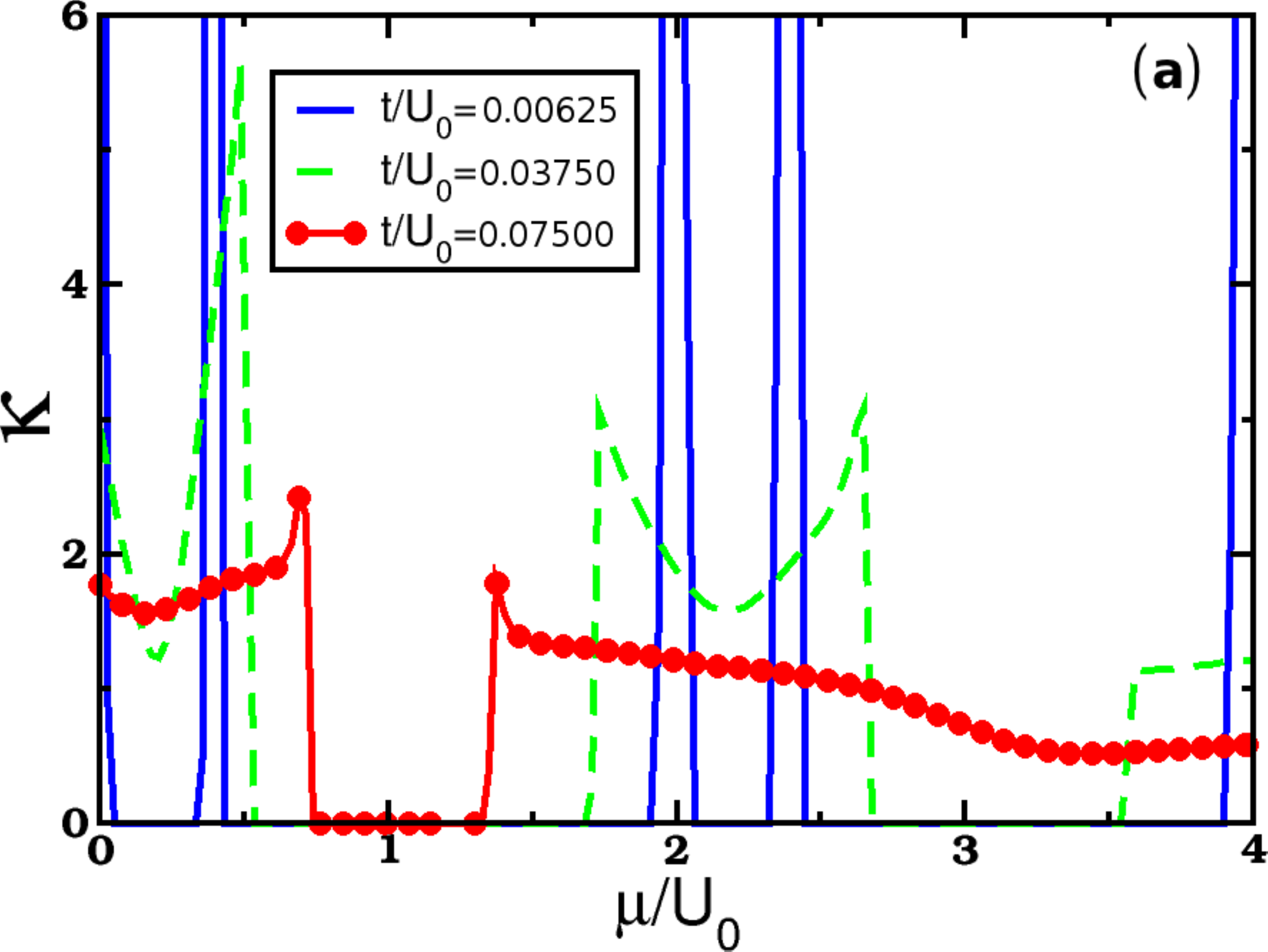}

\includegraphics[scale=0.25]{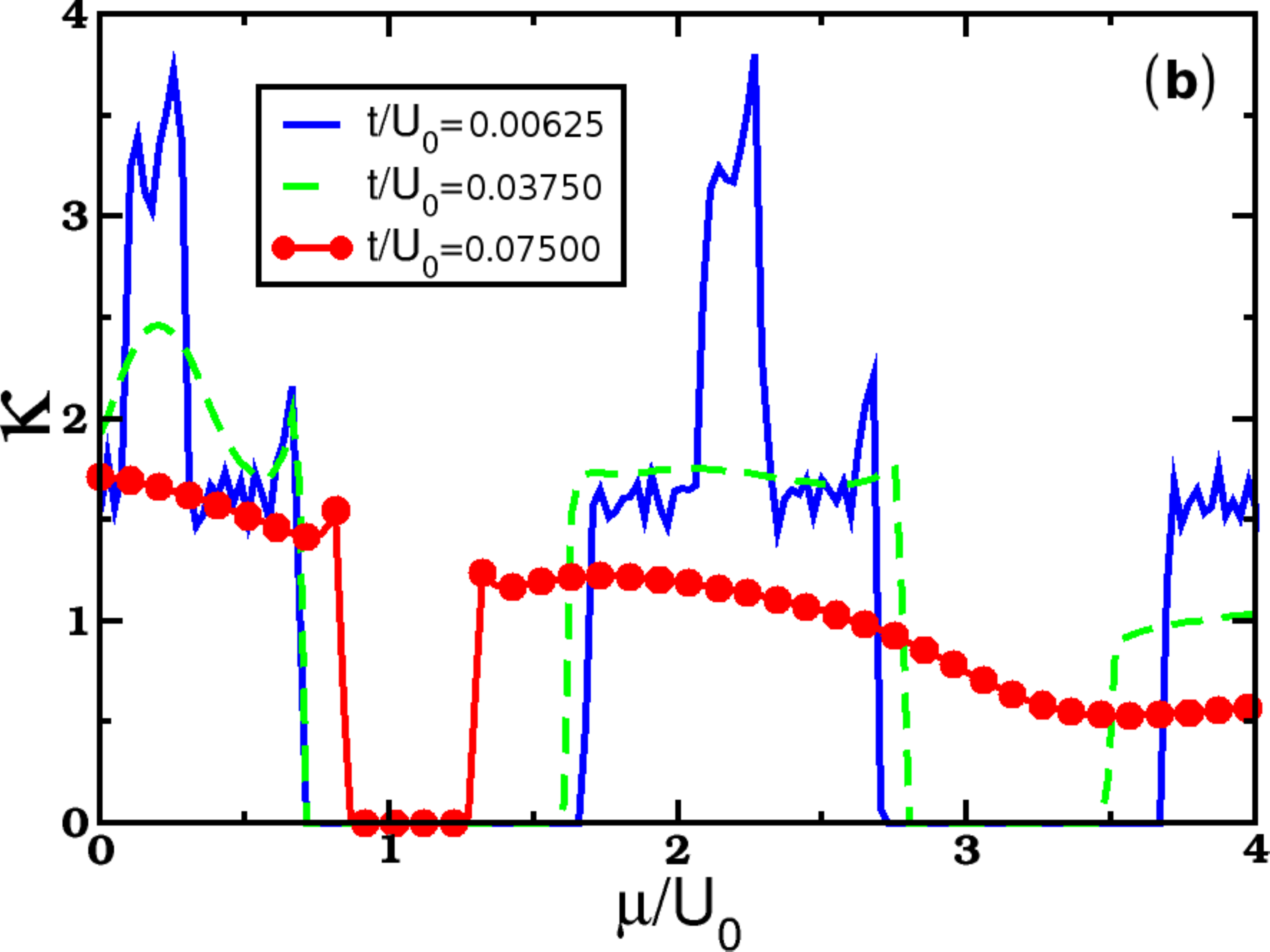} \caption{\label{fig:kappascan}(Color online) The compressibility as a function
of the chemical potential for several values of the hopping amplitude:
(a) clean case and (b) disordered case ($\Delta/U_{0}=0.3$). }
\end{figure}

Fig.~\ref{fig:kappascan} shows some compressibility scans as functions
of $\mu$ for fixed values of $t$; in other words, they correspond
to vertical lines in Fig.~\ref{fig:kappaphasediag}. In the clean
case (Fig.~\ref{fig:kappascan}(a)), the MI regions are clearly demarcated
by the vanishing compressibility. Note the large values of $\kappa$
between MI lobes for $t/U_{0}=0.00625$. Note also that the small odd-numbered
MI lobes can only be seen for this smallest value of hopping amplitude. 

The addition of disorder with strength $\Delta/U_{0}=0.3$ is enough
to completely wipe out the odd-numbered MI lobes, as can be seen in
Fig.~\ref{fig:kappascan}(b). Indeed, it is clear that the compressibility
at $t/U_{0}=0.00625$ (blue curves), which vanishes in extended regions
around $\mu/U_{0}=0.2$ and $2.2$ in Fig.~\ref{fig:kappascan}(a),
becomes non-zero in the same regions after disorder is added, see
Fig.~\ref{fig:kappascan}(b). In fact, it becomes even greater than
in the adjacent regions! For the larger values of hopping shown ($t/U_{0}=0.0375,0.075$),
the system is never in the BG phase (cf. Fig.~\ref{fig:phi1phasediag}(b))
and wherever $\kappa\neq0$ the system is a SF. It is also
noteworthy that the MI lobes that survive have their sizes reduced
when compared with the disorder-free case.

\begin{figure}
\includegraphics[scale=0.25]{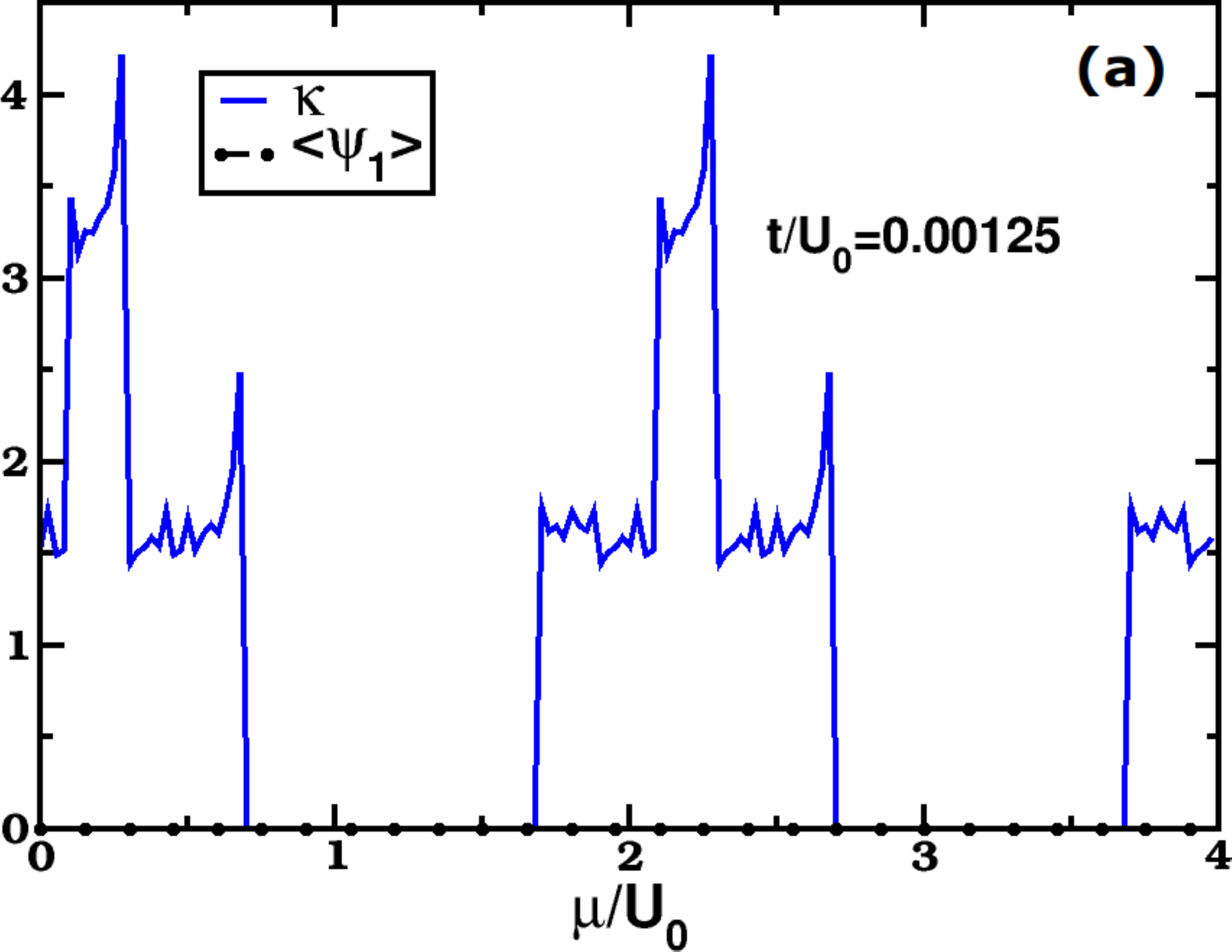}

\includegraphics[scale=0.25]{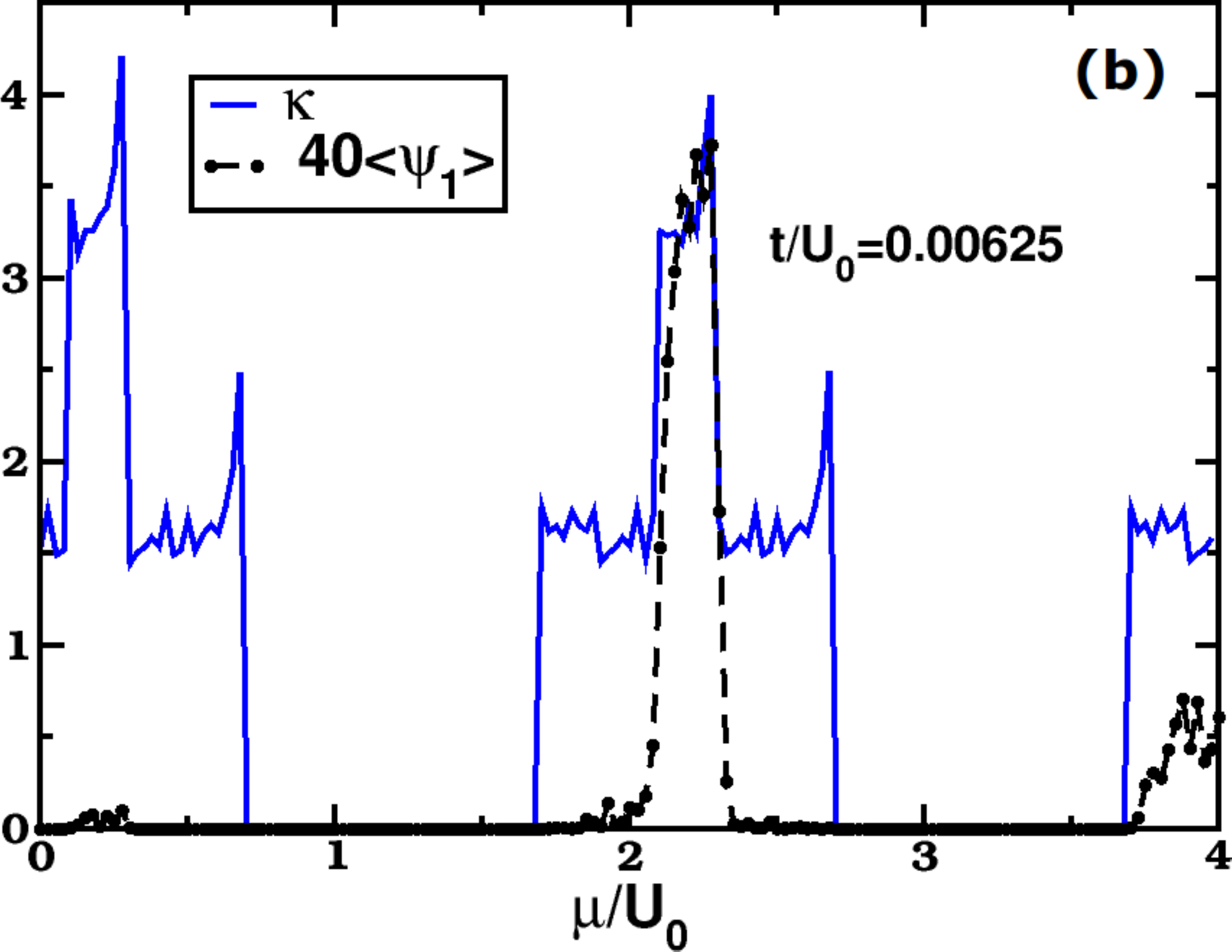} \caption{\label{fig:kappapsi1} (Color online) 
The compressibility $\kappa$
and the average order parameter $\langle\psi_1\rangle$ as functions of the
chemical potential for
(a) $t/U_0=0.00125$ and (b) $t/U_0=0.00625$ ($\Delta/U_{0}=0.3$). }
\end{figure}

In order to further illustrate the joint behavior of the order parameter and the compressibility
for fixed disorder ($\Delta/U_{0}=0.3$), 
we have plotted both quantities together in Fig.~\ref{fig:kappapsi1}  as functions of the chemical potential
for two different values of the hopping amplitude:
(a) $t/U_0=0.00125$ and (b) $t/U_0=0.00625$.
In Fig.~\ref{fig:kappapsi1}(a), the SF phase is never stable and the order parameter vanishes for all values of 
the chemical potential. However, the compressibility is non-zero in large portions of the figure, signaling the BG phase.
In Fig.~\ref{fig:kappapsi1}(b), by contrast, the SF phase emerges out of the regions of enhanced compressibility.
These correspond to the reddish yellow portions of Fig.~\ref{fig:kappaphasediag}(b).

\subsection{Condensate fraction}

\label{sec:c_fraction}

\begin{figure}
\includegraphics[scale=0.3]{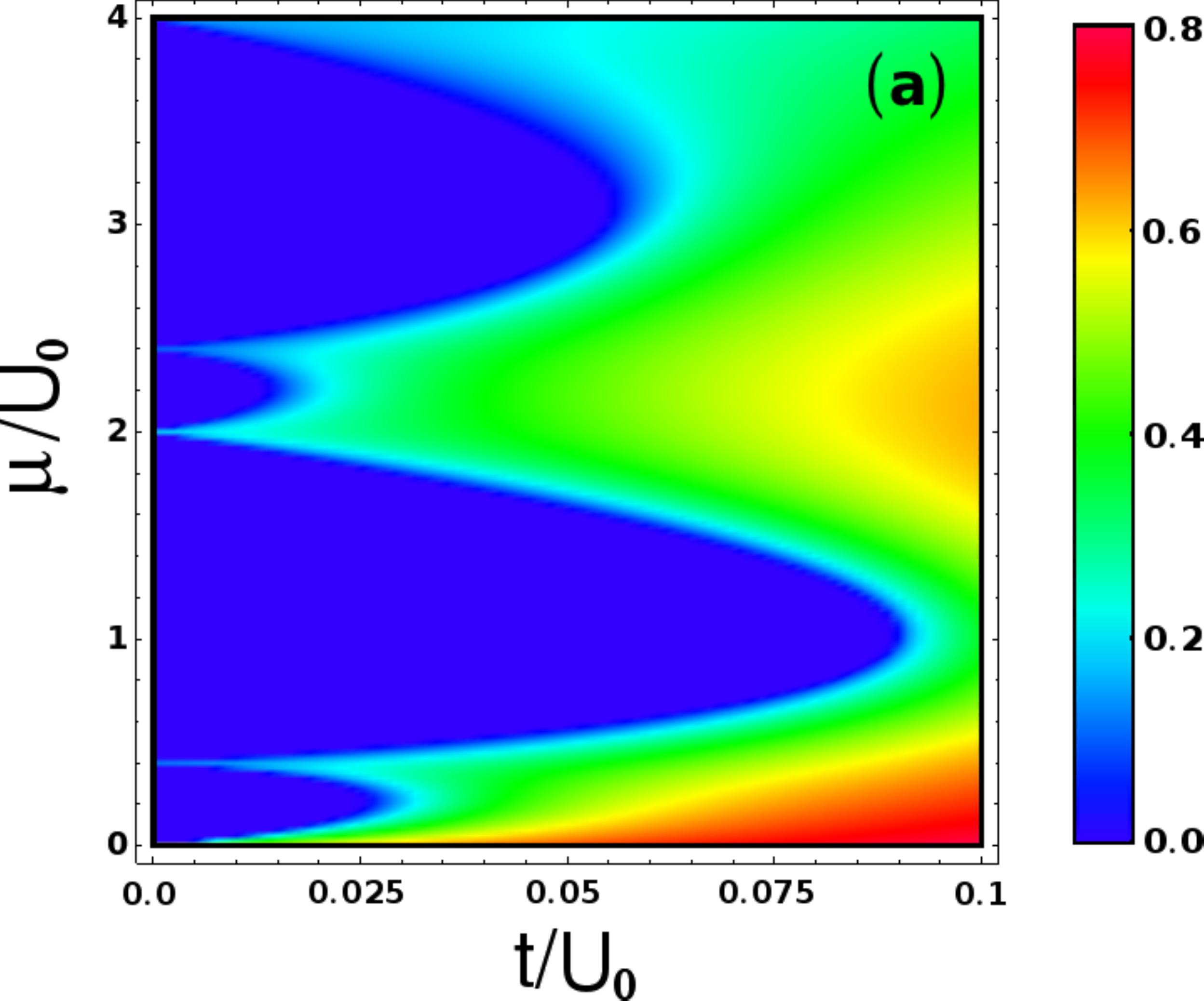}

\includegraphics[scale=0.3]{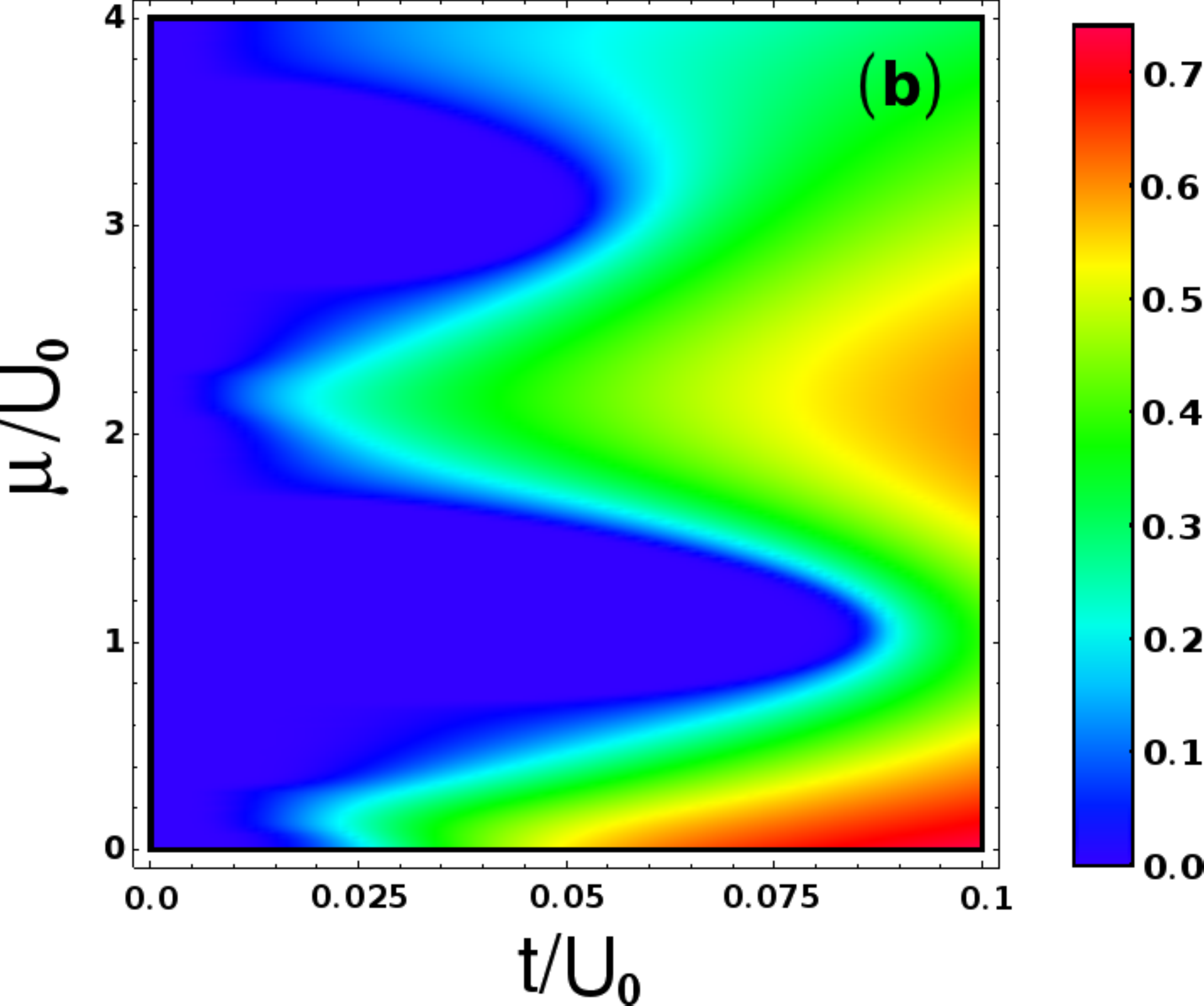} \caption{\label{fig:rhoc}(Color online) The condensate fraction $\rho_{C}$
for the (a) clean and (b) disordered (with $\Delta/U_{0}=0.3$) cases. }
\end{figure}

The condensate fraction within SMFT is given by \cite{BissbortThesis}
\begin{equation}
\rho_{C}=\dfrac{\sum_{\alpha}\left|\psi_{\alpha}\right|^{2}}{n},
\end{equation}
which also serves as an order parameter for the MI-SF phase transition.
This quantity is shown for both the clean and the disordered (with
$\Delta/U_{0}=0.3$) systems in Fig.~\ref{fig:rhoc}. The behavior
in both cases is not qualitatively different from the average order
parameter of Fig.~\ref{fig:phi1phasediag}, as expected.

\subsection{The statistics of the occupation}

\label{sub:occupationstats}

The site occupation number operator $\hat{n}_{i}$ is a very useful
tool for the characterization of the zero-temperature phases of the
clean spin-zero Bose-Hubbard model. Indeed, in the extremely localized
MI limit $t\to0$, the wave function factorizes into uncorrelated
eigenfunctions of $\hat{n}_{i}$ on each site. In this case, the average
occupation $n_{i}=\left\langle \hat{n}_{i}\right\rangle $ equals
one of the integer eigenvalues and quantum fluctuations of the occupation,
as measured by the standard deviation 
\begin{equation}
\Delta n_{i}=\sqrt{\left\langle \hat{n}_{i}^{2}\right\rangle -\left\langle \hat{n}_{i}\right\rangle ^{2}},\label{eq:occupationstdev}
\end{equation}
are evidently zero. On the other hand, in the other extreme limit
of a weakly correlated SF $U\to0$, the site occupation number operator
is not a good quantum number and there are large quantum fluctuations
signaled by a non-zero $\Delta n_{i}$. In the clean case, lattice
translation invariance guarantees that both $n_{i}$ and $\Delta n_{i}$
are uniform and do not depend on the site $i$. Once disorder is added,
however, spatial fluctuations of both quantities arise, in addition
to the quantum fluctuations already present in the clean system.

A useful measure of these fluctuations is afforded by the distribution
function $P_{n}\left(n_{i}\right)$ and $P_{\Delta n}\left(\Delta n_{i}\right)$,
which are both very easily obtained within SMFT from the solutions
of the ensemble of single-site Hamiltonians of Eq.~\eqref{eq:singlesiteham}.
We will thus now show our results for these distributions for the
disordered spin-1 Bose-Hubbard model.

We start by looking at the spatial average of $n_{i}$, which gives
the average number of bosons per site $n$. The two figures can hardly
be distinguished, although tiny distortions can be seen. The value
of $n$ is useful if we want to assign an integer to the MI lobe,
but it is not very useful for a precise demarkation of the phases.
However, as we will see, in contrast to the average $n$ the full
statistics of $n_{i}$ imparts a great deal of useful information.

\begin{figure}
\includegraphics[scale=0.3]{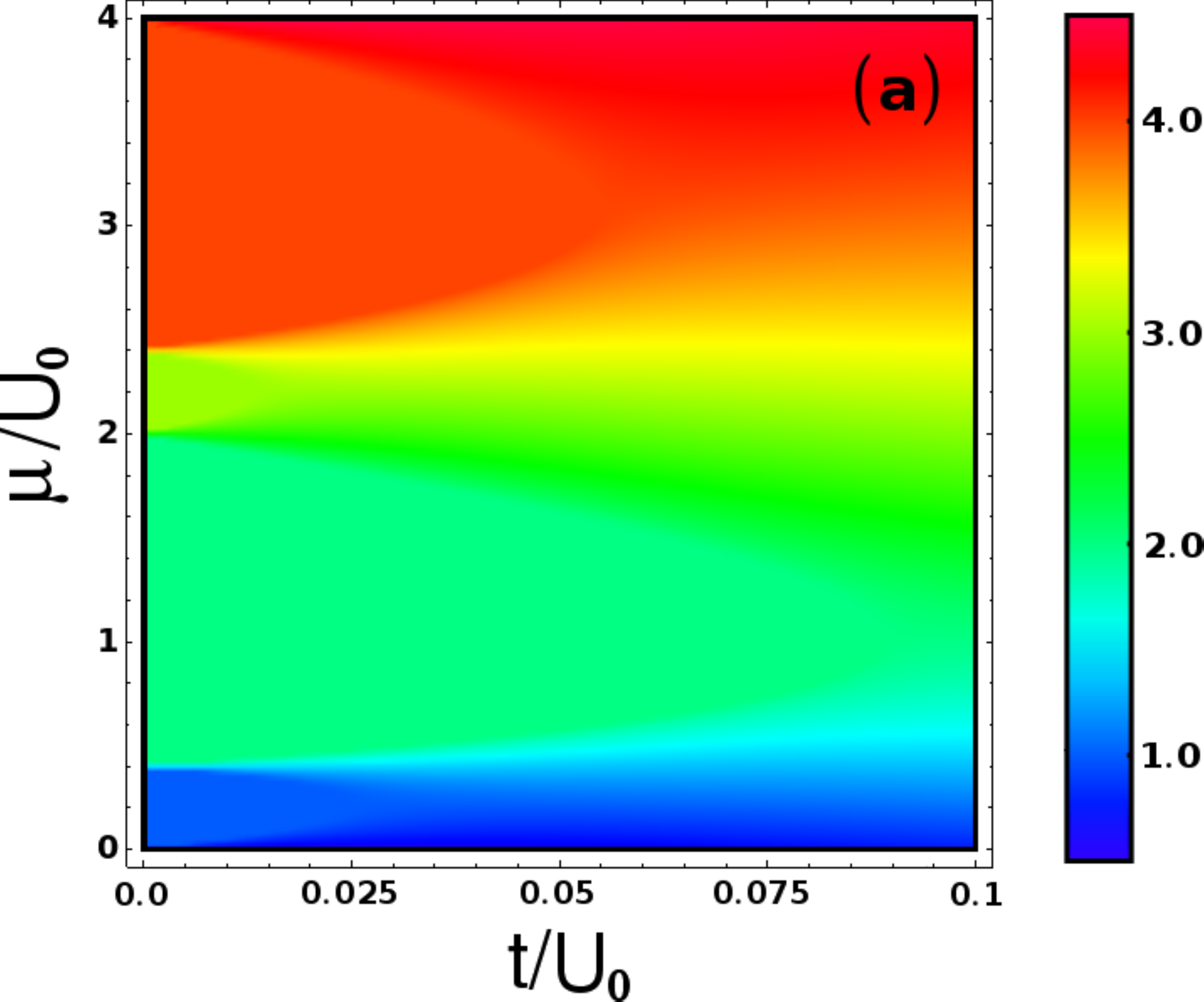}

\includegraphics[scale=0.3]{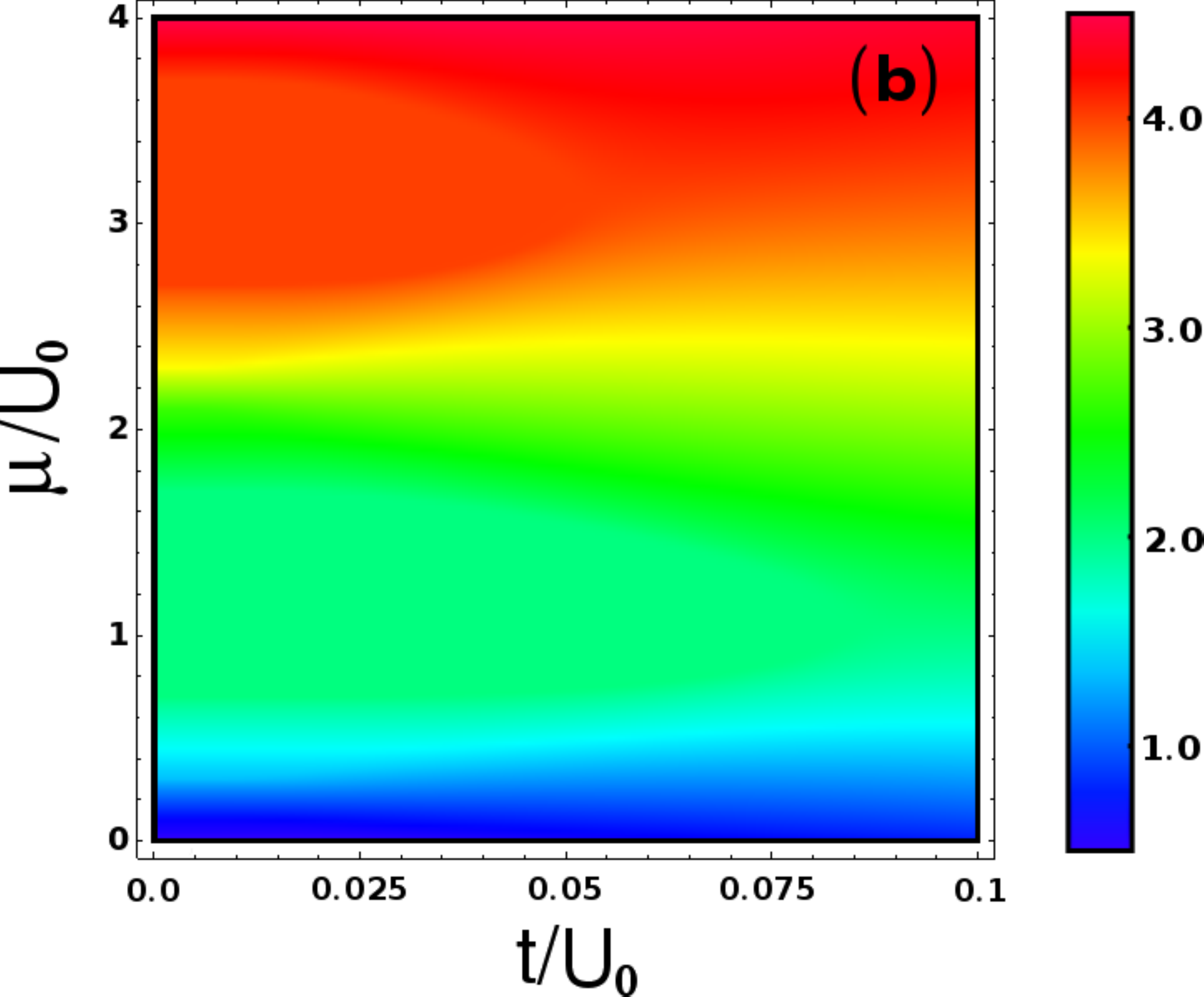} \caption{\label{fig:averagen}(Color online) The average site occupation number
$n$ for the (a) clean and (b) disordered ($\Delta/U_{0}=0.3$) cases
in the $\mu$ vs $t$ plane.}
\end{figure}

\begin{figure}
\includegraphics[scale=0.25]{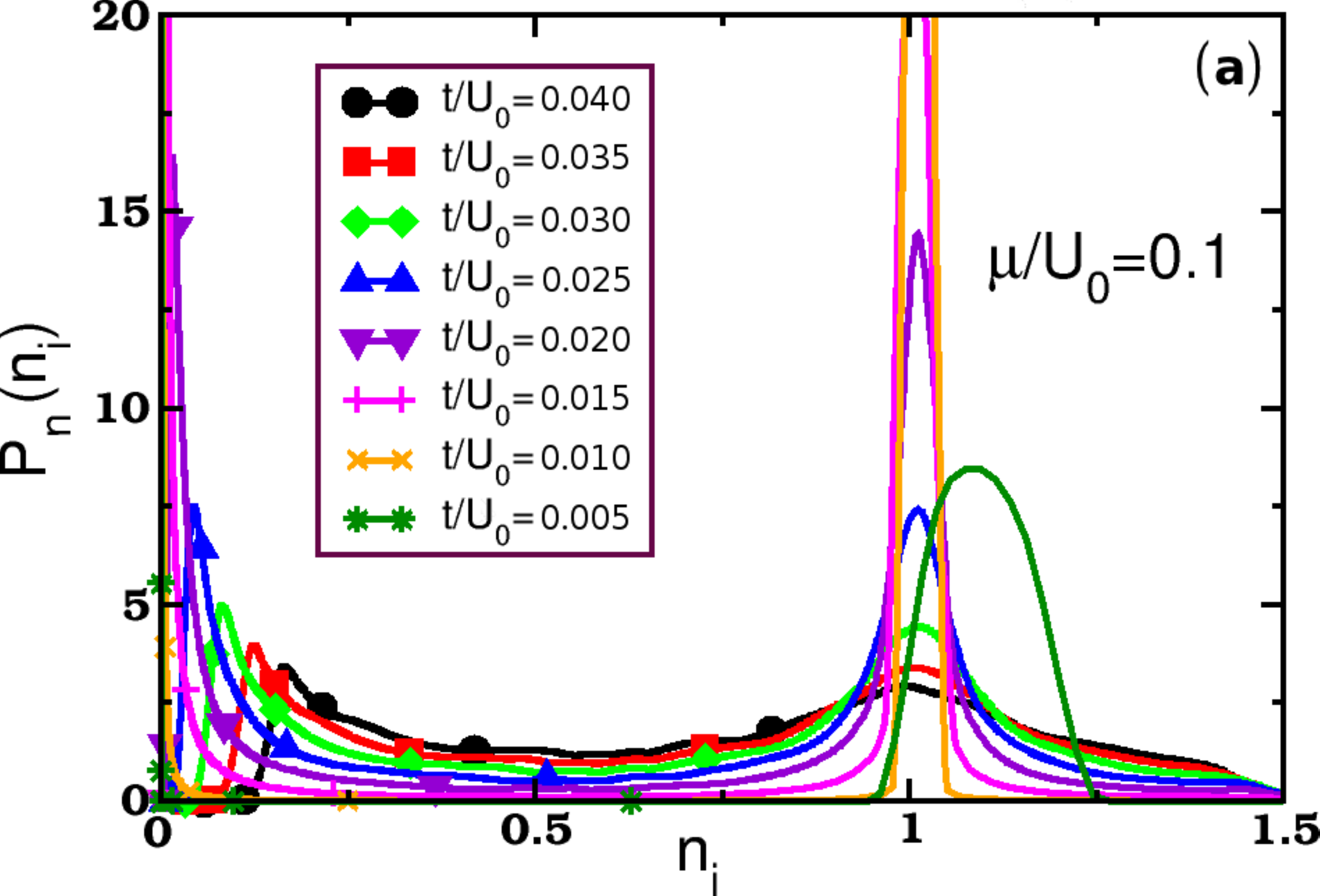}

\includegraphics[scale=0.25]{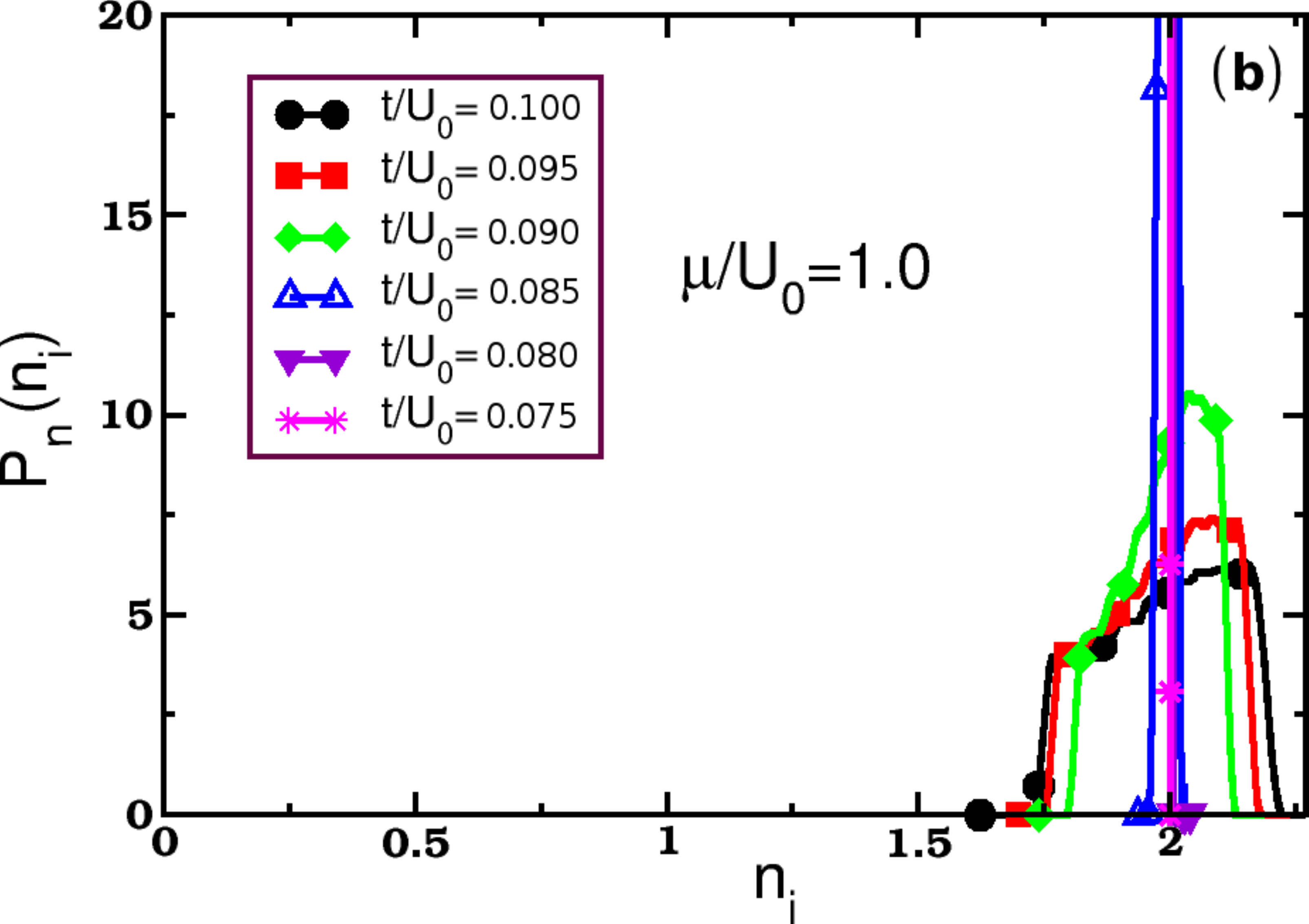} \caption{\label{fig:p_n}(Color online) The probability distribution functions
of the average site occupation number $P_{n}(n_{i})$ for (a) $\mu/U_{0}=0.1$,
and (b) $\mu/U_{0}=1$ and various values of the hopping amplitude.
In (a) the system goes through the SF-BG transition and in (b) from
SF to MI, as the hopping amplitude decreases. The disorder is set
to $\Delta/U_{0}=0.3$.}
\end{figure}

In Fig.~\ref{fig:p_n} $P_{n}(n_{i})$ is shown for two values of
chemical potential, $\mu/U_{0}=0.1$ (Fig.~\ref{fig:p_n}(a)) and
$\mu/U_{0}=1$ (Fig.~\ref{fig:p_n}(b)), and for various values of
$t/U_{0}$. In Fig.~\ref{fig:p_n}(a), the system goes from a SF
to a BG as the hopping decreases. The spatial fluctuations of $n_{i}$
are large in both phases. As the hopping decreases and the system
approaches the BG phase, the distribution function acquires a bimodal
shape, with increasingly sharper peaks around $n_{i}=0$ and $n_{i}=1$
and decreasing weight in the region between these two values. Inside
the BG phase ($t/U_{0}=0.01$ and $0.005$), the peak around $n_{i}=0$
becomes a delta function while the peak at $n_{i}=1$ remains broad.
This spatial landscape in which different sites are Mott localized
at different occupations is characteristic of the BG phase \cite{fisher89}
and is vividly illustrated by $P_{n}\left(n_{i}\right)$.

The behavior observed across the SF-MI phase transition is markedly
different, as can be seen in Fig.~\ref{fig:p_n}(b). In this case,
$P_{n}\left(n_{i}\right)$ starts as a mildly broad distribution around
$n_{i}=2$ in the SF, which becomes increasingly narrower as the hopping
is reduced and the system transitions into the MI phase. Inside the
MI lobe ($t/U_{0}=0.08$ and $0.075$), the distribution becomes a
delta function centered at $n_{i}=2$, showing that in the disordered
MI the system is locked at a fixed unique occupation.

\begin{figure}
\includegraphics[scale=0.3]{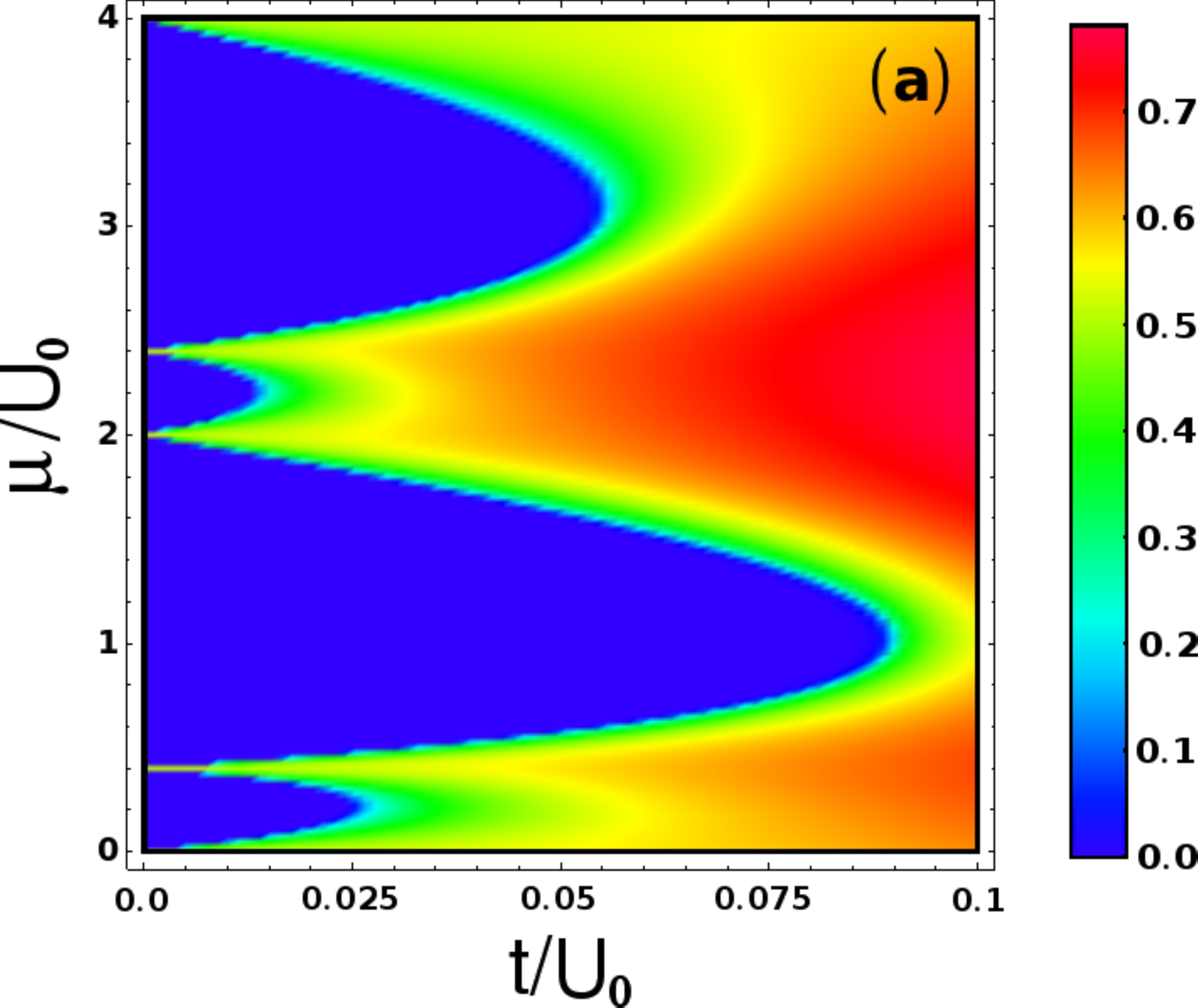}

\includegraphics[scale=0.3]{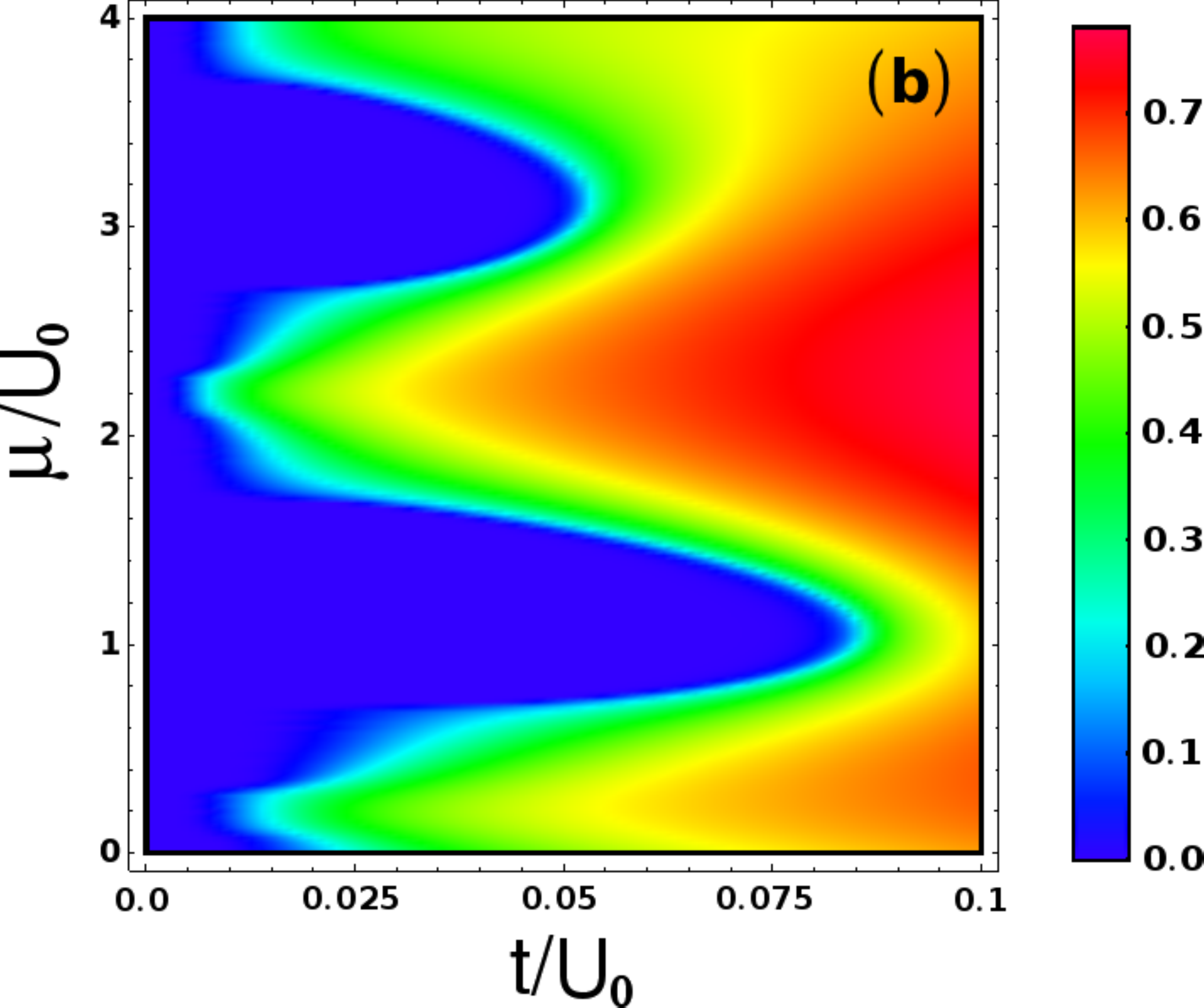} \caption{\label{fig:deltana}(Color online) Spatial average of the site occupation
standard deviation $\Delta n_{i}$ in the $\mu$ vs $t$ plane: (a)
clean and (b) disordered ($\Delta/U_{0}=0.3$) cases. }
\end{figure}

We now turn to the spatial fluctuations of $\Delta n_{i}$ as measured
by $P_{\Delta n}\left(\Delta n_{i}\right)$. We start by looking at
the spatial average of $\Delta n_{i}$ in Fig. ~\ref{fig:deltana}.
Within SMFT, both MI and BG phases are characterized by the vanishing
of the order parameter $\psi_{i\alpha}$ and thus of the $\eta_{i\alpha}$
field that acts on each site, see Eqs.~\eqref{eq:etadefinition}
and \eqref{eq:singlesiteham}. If $\eta_{i\alpha}$ is zero, the ground
state of every site an eigenvector of the number operator and, therefore,
$\Delta n_{i}=0$ for all sites. This is why the average $\Delta n_{i}$
is also zero within both the MI and the BG phases. This is a feature
of the mean field character of the theory and is not expected to survive
beyond this approximation. In contrast, $\Delta n_{i}\neq0$ everywhere
in the SF phase and so is its average, making it an alternative order
parameter for that phase in the clean as well as in the disordered
cases. Note that, although in the BG phase $\Delta n_{i}=0$ at every
site and there are no quantum fluctuations of the site occupation
(within SMFT), the average site occupation does exhibit large spatial
fluctuations, as was already seen in Fig.~\eqref{fig:p_n}(a).

\begin{figure}
\includegraphics[scale=0.25]{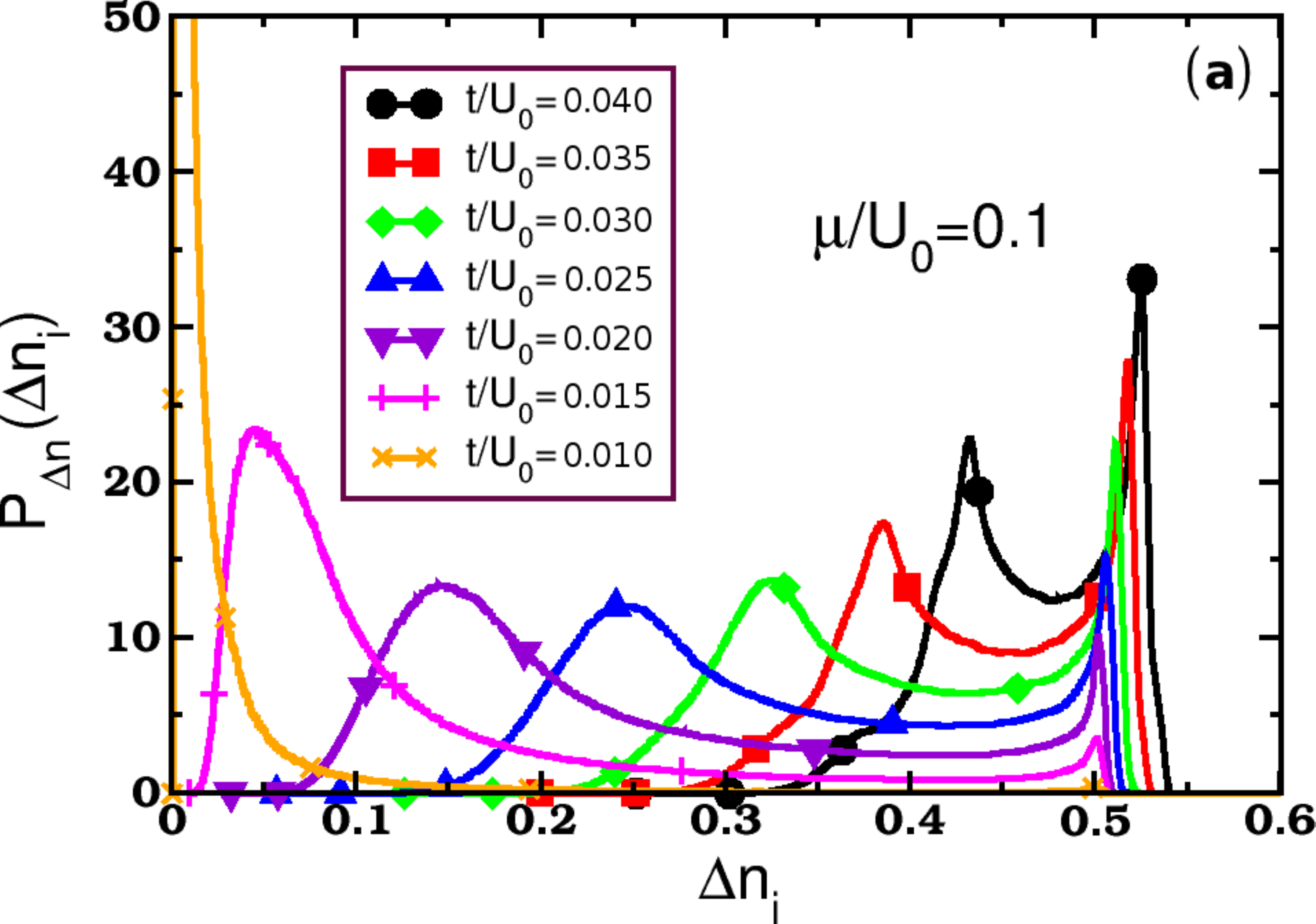}

\includegraphics[scale=0.25]{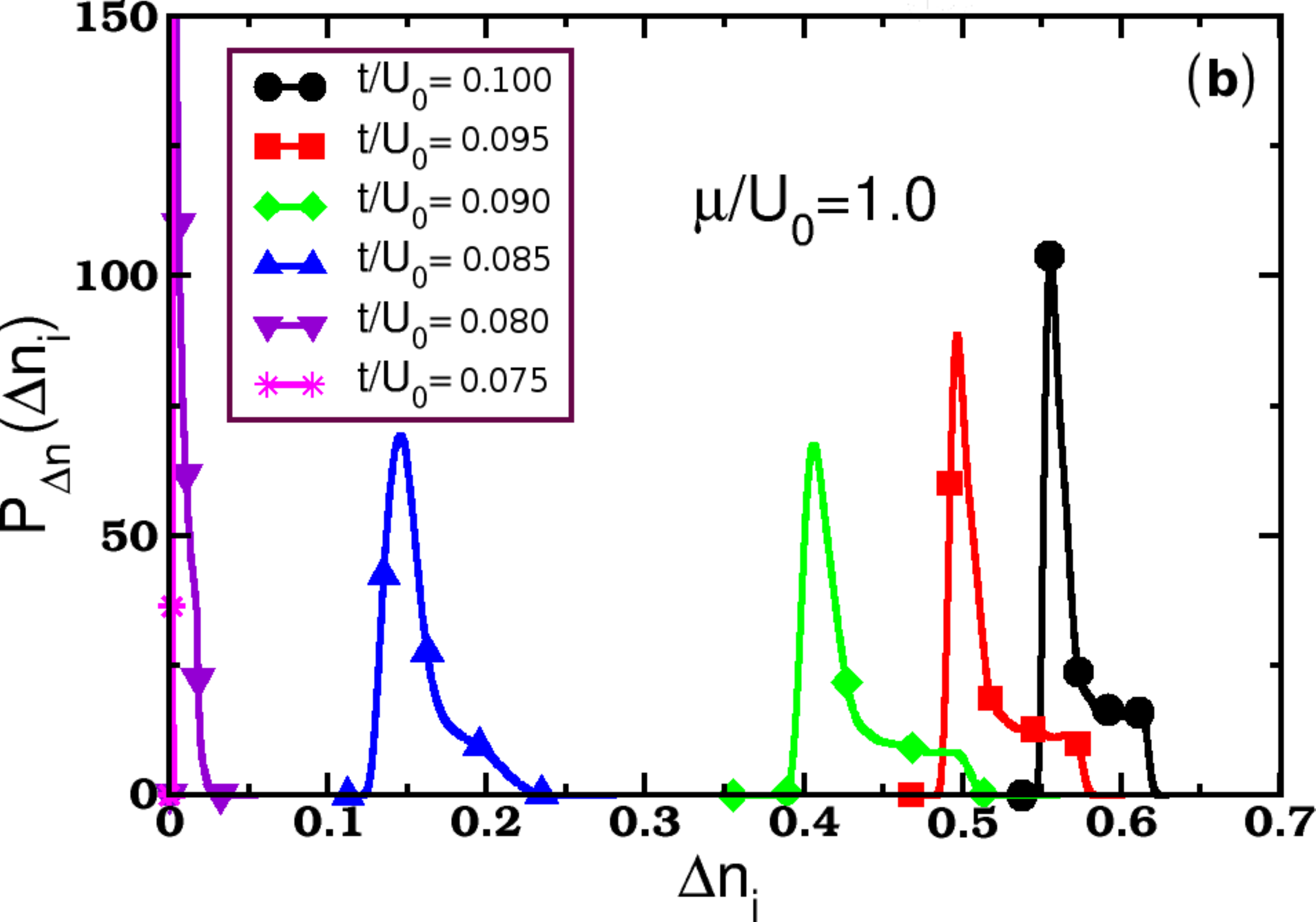} \caption{\label{fig:deltanb}(Color online) Probability distributions of the
site occupation standard deviation $\Delta n_{i}$ for two values
of the chemical potential: (a) $\mu/U_{0}=0.1$ (corresponding to
the SF to BG transition) and (b) $\mu/U_{0}=1.0$ (which corresponds
to the transition from SF to MI), for various values of the hopping
amplitude and for fixed disorder strength $\Delta/U_{0}=0.3$. }
\end{figure}

The full distributions $P_{\Delta n}\left(\Delta n_{i}\right)$ are
shown in Fig.~\ref{fig:deltanb} for the two chemical potential values
$\mu/U_{0}=0.1$ and $1.0$ that allow us to study the SF to BG and
MI phase transitions. In the first case (Fig.~\ref{fig:deltanb}(a)),
the distribution is mildly broad, approximately bimodal and with support
around $\Delta n_{i}\approx0.5$ in the SF. As $t$ decreases and
the system approaches the BG, the distribution widens with a small
and sharp peak at $\approx0.5$ and a broader one centered at a lower
value which slowly shifts towards zero while at the same time gaining
more weight. Eventually, in the BG phase, the distribution degenerates
into a delta function at zero, consistent with the vanishing average
value found before.

Finally, the behavior of $P_{\Delta n}\left(\Delta n_{i}\right)$
for the SF to MI transition case in shown in Fig.~\ref{fig:deltanb}(b).
The behavior is now distinctively different: the distributions are
always confined to a small region of support, whose center shifts
towards zero and whose width decreases as the system enters the MI
phase. As discussed before, the presence of local occupation number
quantum fluctuations is intimately tied to the non-zero value of $\eta_{i\alpha}$.
Therefore, it should not be viewed as too surprising that the qualitative
behavior of $P_{\Delta n}\left(\Delta n_{i}\right)$ closely follows
that of $P_{1}\left(\psi_{1}\right)$, cf. Figs.~\ref{fig:phi1dist}
and \ref{fig:deltanb}.

\subsection{Spin}

\label{sec:spin}

\begin{figure}
\includegraphics[scale=0.3]{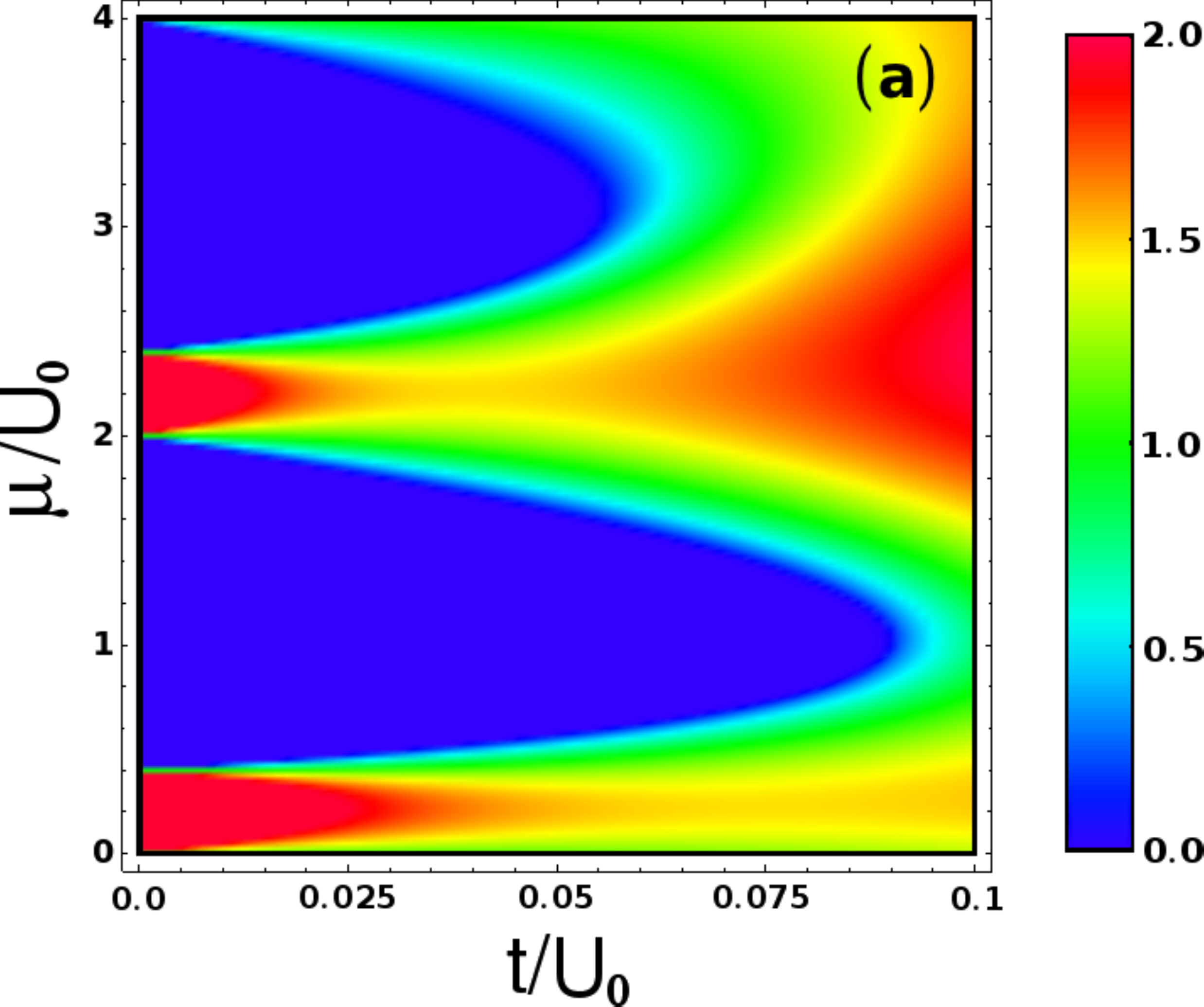}

\includegraphics[scale=0.3]{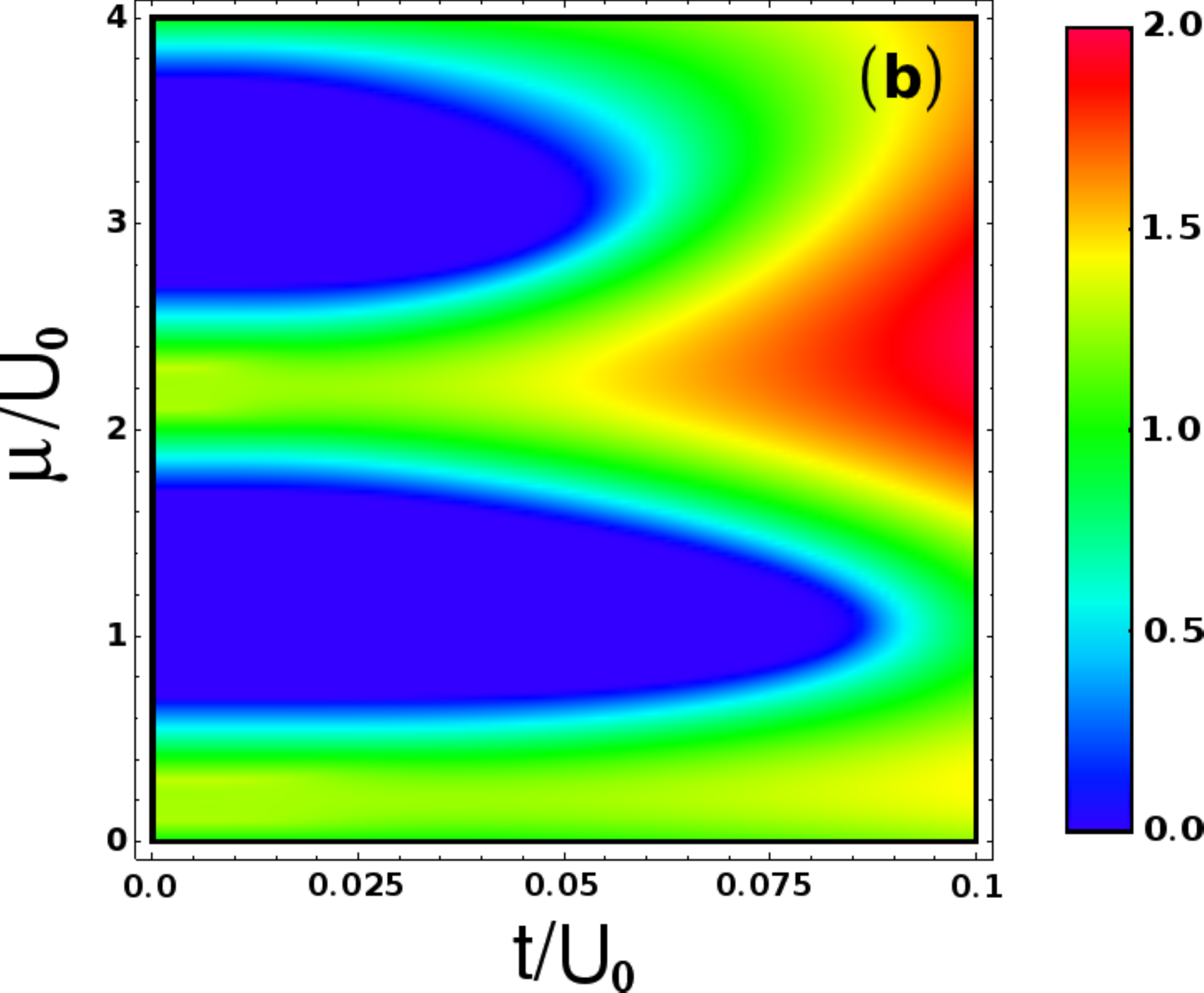} \caption{\label{fig:ss_a}(Color online) The spatial average of the mean square
of the total spin per site $\langle\boldsymbol{S}_{i}^{2}\rangle$:
(a) clean and (b) disordered ($\Delta/U_{0}=0.3$) cases. }
\end{figure}

Another quantity of importance in the characterization of the phases
is the average square of the total spin of each site $\left\langle \boldsymbol{S}_{i}^{2}\right\rangle \equiv S_{i}^{2}$.
In the clean limit, this quantity is zero in the even-numbered MI
lobes, since the bosons are able to combine into a zero-spin composite
at each site thus decreasing the spin-dependent interaction contribution
to the total energy. In contrast, it is impossible to do so when there
is an odd number of bosons per site and the best compromise to lower
the energy is to form a spin-1 combination, in which case $S_{i}^{2}=2$.
This situation is depicted in Fig.~\ref{fig:ss_a}(a). Interestingly,
the polar SF is characterized in general by intermediate values of
this quantity, with a tendency towards saturation to $S_{i}^{2}=2$
when the hopping is large and the SF well formed. It should be noted
that inter-site spin correlations, absent in the mean-field treatment
used here, are able to generate complex spin arrangements in the ground
state. In particular, spin nematic order is predicted to occur throughout
the odd-numbered MI lobes and in part of the even-numbered ones \cite{imambekov2003}.
This type of order is characterized by broken spin rotational invariance
($\left\langle \left(S_{i}^{a}\right)^{2}\right\rangle \neq0$, $a=x,y$
or $z$) accompanied by unbroken time reversal symmetry ($\left\langle \boldsymbol{S}_{i}\right\rangle =0$).

The most dramatic effect of the introduction of disorder is seen in
the BG phase, see Fig.~\ref{fig:ss_a}(b). In that case, the presence
of sites with different average occupations, both even and odd (see
Fig.~\ref{fig:p_n}(a)), leads to the settling of the spatial average
of $S_{i}^{2}$ at a value intermediate between 0 and 2. There is
actually a very smooth dependence of this spatial average on $t$
as we move from the SF into the BG phase. In the MI phase, on the
other hand, the spatial average of $S_{i}^{2}$ still vanishes and
in the SF it also retains its generic intermediate values.

\begin{figure}
\includegraphics[scale=0.24]{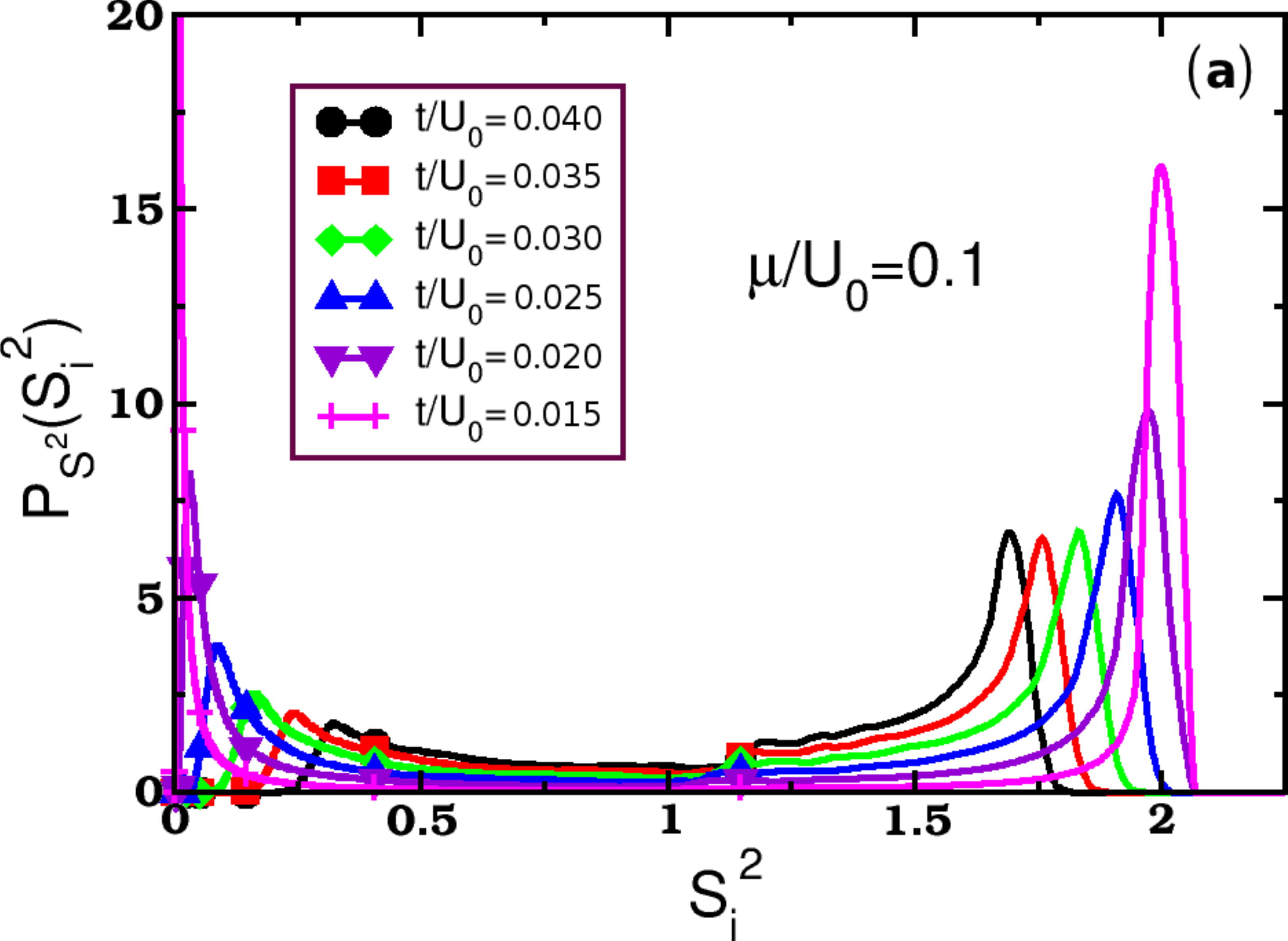}

\includegraphics[scale=0.24]{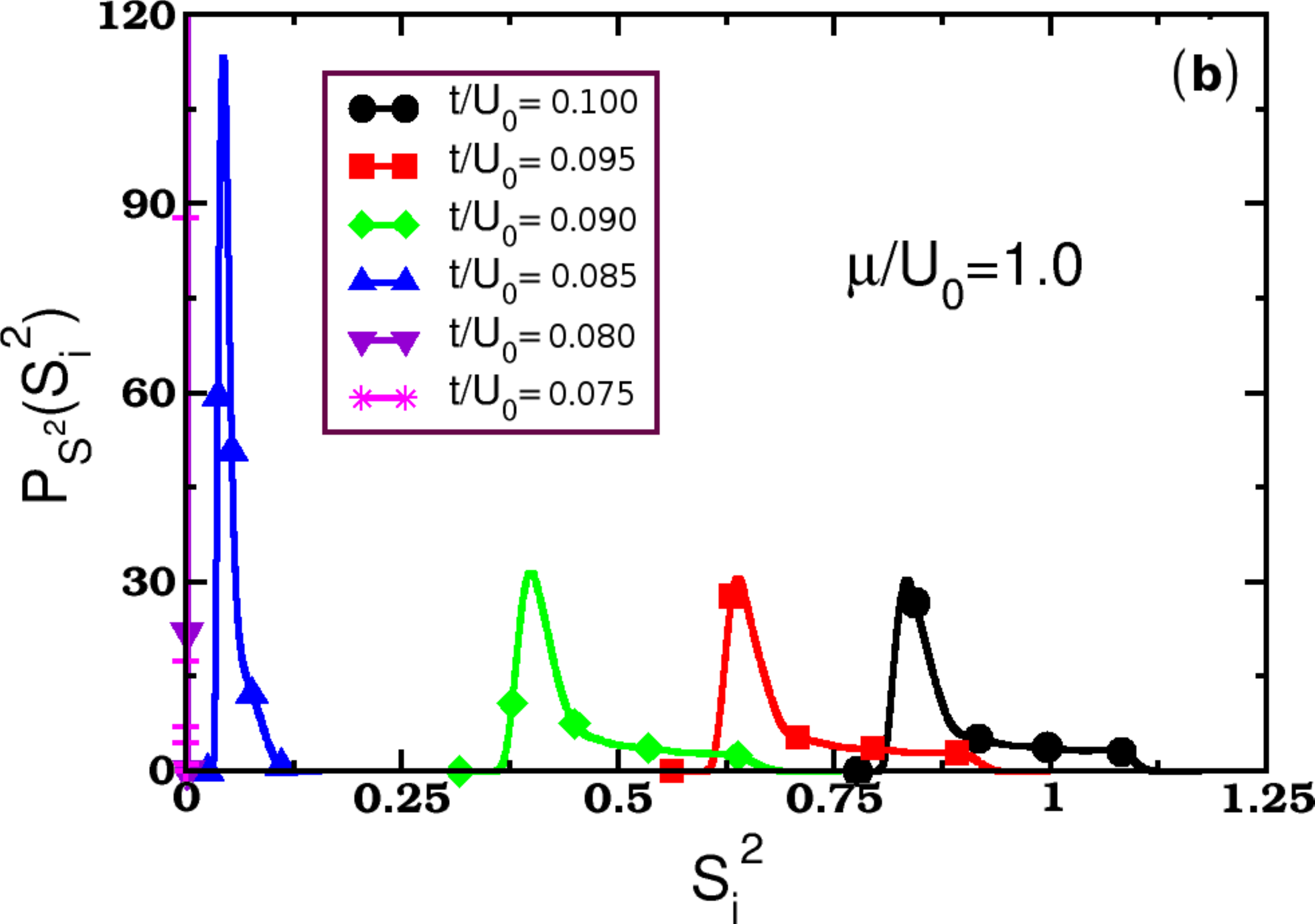} \caption{\label{fig:ss_b}(Color online) The probability distribution functions
$P_{S^{2}}\left(S_{i}^{2}\right)$ of the square of the total spin
per site for various values of the hopping amplitude and two values
of the chemical potential: (a) $\mu/U_{0}=0.1$ and (b) $\mu/U_{0}=1.0$.
The disorder is set to $\Delta/U_{0}=0.3$.}
\end{figure}

The probability distribution of the expectation value of the square
of the total spin per site $P_{S^{2}}(S_{i}^{2})$ is shown in Fig.~\ref{fig:ss_b}
for the two values of $\mu/U_{0}=0.1$ and $1.0$ and several values
of the hopping amplitude. From the previous discussion, the behavior
of $P_{S^{2}}(S_{i}^{2})$ is expected to track closely the distribution
of the average site occupation $P_{n}\left(n_{i}\right)$. Indeed,
upon approaching the BG from the SF as $t$ decreases, as shown in
Fig.~\ref{fig:ss_b}(a), the distribution of $S_{i}^{2}$ becomes
increasing broader with a bimodal shape, indicating the gradual appearance
of both spin-zero and spin-1 sites, corresponding to the peaks at
$n_{i}=0$ and $n_{i}=1$, respectively, of Fig.~\ref{fig:p_n}(a).
Likewise, as $t$ decreases and the system transitions from the SF
to the MI, as depicted in Fig.~\ref{fig:ss_b}(b), the $S_{i}^{2}$
distribution shifts weight from non-zero values spread around $\approx1$
down to a delta function at zero, at the same time as the average
occupation distribution narrows down to a delta function at occupation
$n_{i}=2$.

It was argued in reference \onlinecite{sanpera11} that the \emph{spatially-averaged}
value of $\langle\boldsymbol{S}_{i}^{2}\rangle$ intermediate between
0 and 2 of the BG phase shown in Fig.~\ref{fig:ss_a}(b) is indicative
of a spin nematic phase \cite{imambekov2003}. In the clean case discussed
in \onlinecite{imambekov2003}, however, it is quantum inter-site spin correlations
that are responsible for the appearance of nematic order. This occurs
even at perturbatively small $t$, in which case each site has a fixed
odd number of bosons. On the other hand, no inter-site spin correlations
are incorporated in the SMFT and the BG phase is characterized by
the presence of sites with different number occupations, see Fig.~\ref{fig:p_n}(a).
It is these sites, with an odd number of bosons and $S_{i}=1$ or
an even number and $S_{i}=0$, which are ultimately responsible for
the intermediate value of $\overline{\langle\boldsymbol{S}_{i}^{2}\rangle}$
which both the SMFT and the Gutzwiller approach of reference \onlinecite{sanpera11}
find. This situation is rather different from the clean nematic and
should not, in our view, be confused with it. 

\subsection{The quantum phase transition as a function of disorder}

\label{sec:disorder_function}

In previous Sections we fixed the disordered strength and analyzed
the behavior of the system as a function of the chemical potential
and the hopping amplitude. For the value of disorder we used ($\Delta/U_{0}=0.3$),
the clean even-numbered MI lobes survived the introduction of randomness,
whereas the odd-numbered ones were completely destroyed. It would
be interesting to see how the former behave as the disorder strength
is further increased. We take up this task in this Section.

\begin{figure}
\includegraphics[scale=0.25]{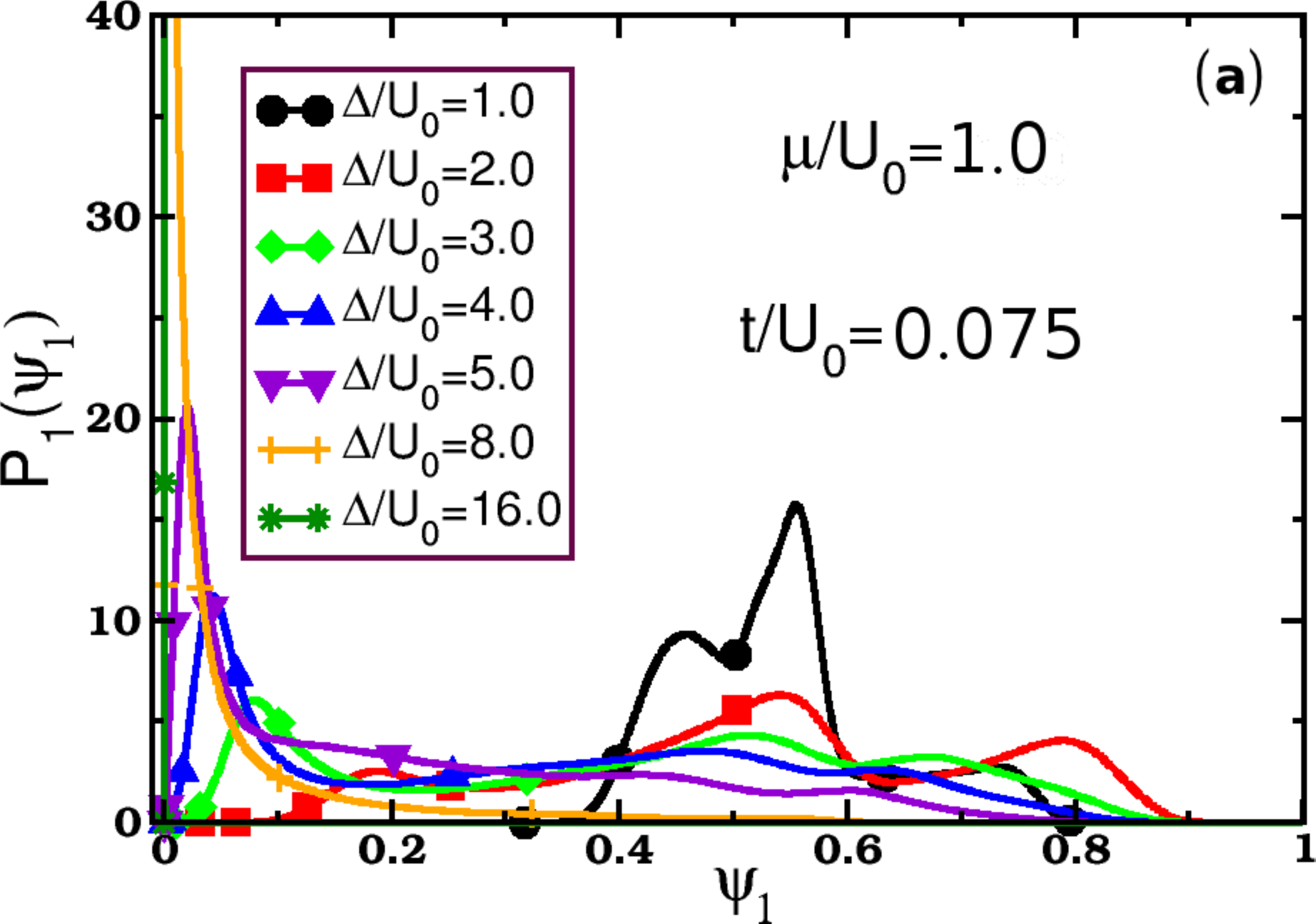}

\includegraphics[scale=0.25]{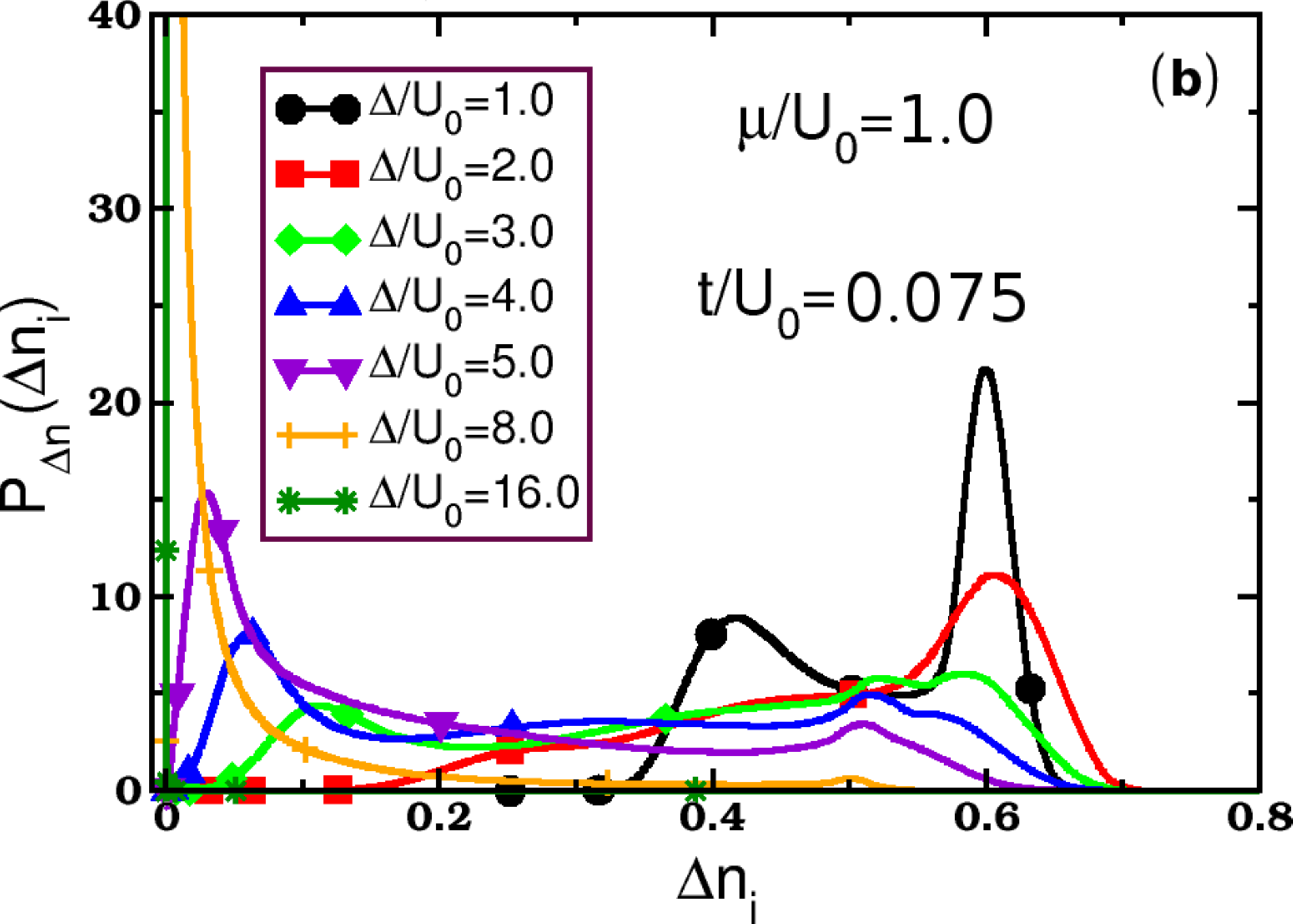} \caption{\label{fig:s1_todosa}(Color online) Probability distributions functions
of: (a) the order parameter and (b) the standard deviation of the
occupation, for various values of disorder parameter $\Delta/U_{0}$.
The chemical potential is fixed at $\mu/U_{0}=1$ and the hopping
amplitude at $t/U_{0}=0.075$.}
\end{figure}

We show in Fig.~\ref{fig:s1_todosa} the various distribution functions
in the strong disorder regime $(\Delta\geq U_{0})$ for $\mu/U_{0}=1.0$
and $t/U_{0}=0.075$. We remind the reader that for $\Delta=0$, this
corresponds to a point well inside the $n_{i}=2$, $S_{i}=0$ MI lobe.
The order parameter distribution $P_{1}\left(\psi_{1}\right)$ is
shown in Fig.~\ref{fig:s1_todosa}(a). For $\Delta/U_{0}=1.0$ the
MI lobe has been suppressed in favor of the disordered SF phase, characterized
by finite SF order parameters. As $\Delta$ is increased, this distribution
broadens with increasing weight at $\psi_{1}=0$. Eventually, for
large enough randomness, the distribution collapses to a delta function
at $\psi_{1}=0$, signaling the destruction of the SF phase. The distribution
of site occupation standard deviation $P_{\Delta n}\left(\Delta n_{i}\right)$,
depicted in Fig.~\ref{fig:s1_todosa}(b), shows a qualitatively similar
behavior, as expected. As discussed in Section \ref{sub:occupationstats},
this quantity largely tracks the distribution of the order parameters.
But is the non-SF phase at large values of $\Delta$ a BG or a MI?

\begin{figure}
\includegraphics[scale=0.25]{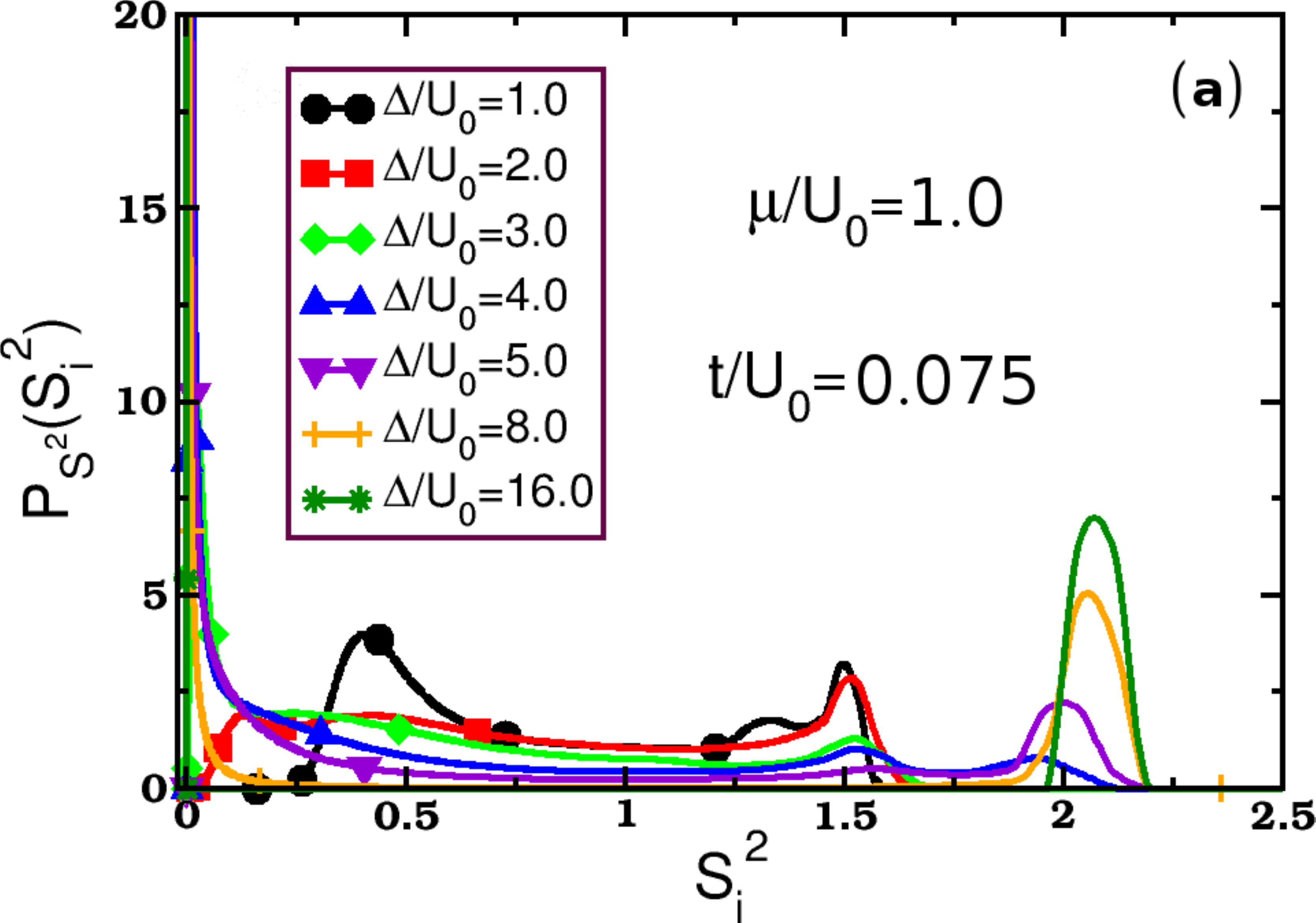}

\includegraphics[scale=0.25]{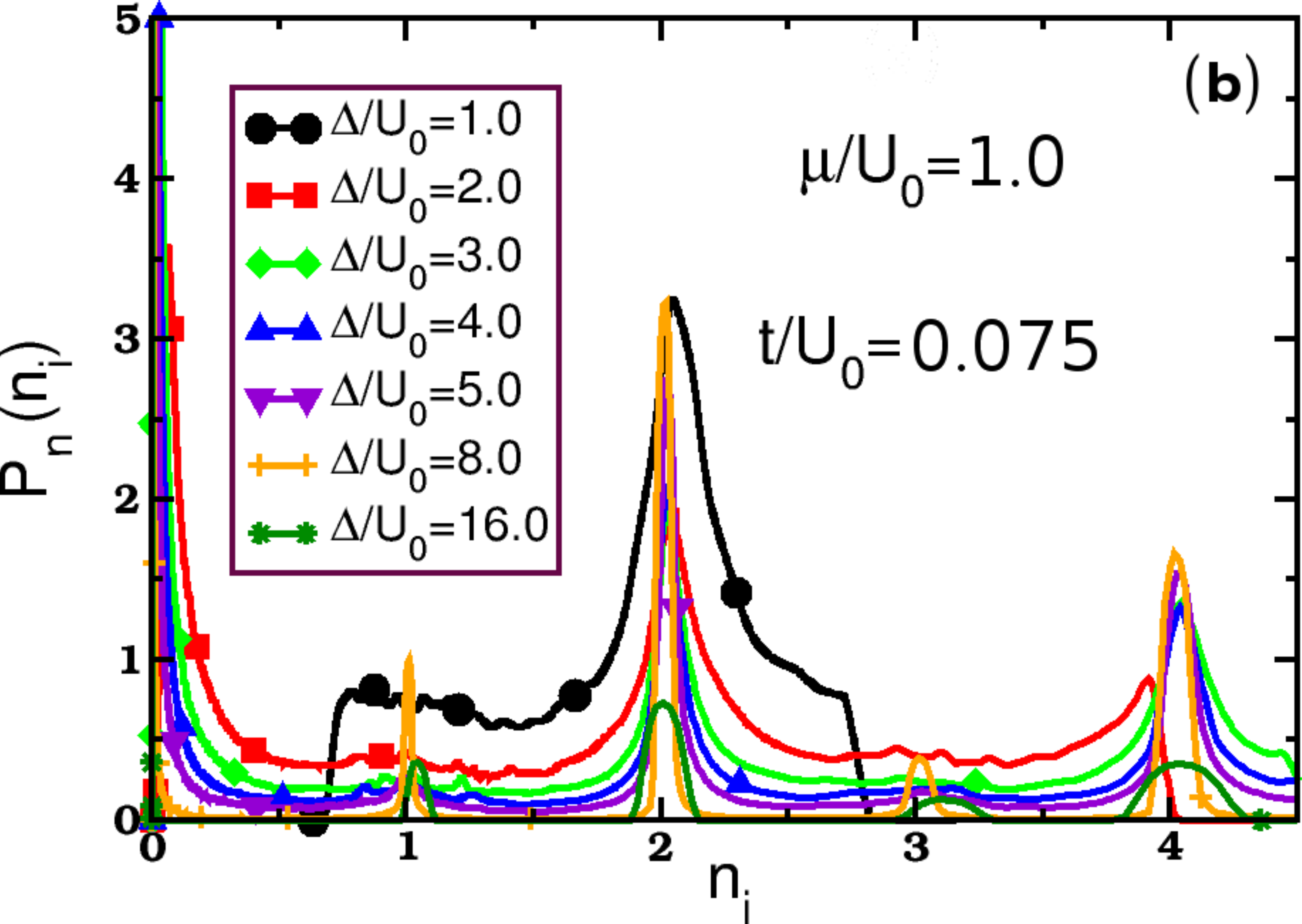} \caption{\label{fig:s1_todosb}(Color online) Probability distributions functions
of: (a) the expectation value of the square of the spin operator per
site and (b) the average site occupation number, for various values
of disorder parameter $\Delta/U_{0}$. The chemical potential is fixed
at $\mu/U_{0}=1$ and the hopping amplitude at $t/U_{0}=0.075$.}
\end{figure}

One possible diagnostics tool is afforded by the distribution of the
mean square of the total spin per site, as shown in Fig.~\ref{fig:s1_todosb}(a).
It shows the typical broad, bimodal shape characteristic of the BG
distribution as $\Delta$ increases, indicating the presence of both
singlet (weight at 0) and spin-1 (weight around 2) composites at each
site. This should be compared with the similar small $t$ distributions
of Fig.~\ref{fig:ss_b}(a), characteristic of the BG phase, and contrasted
with the corresponding curves of Fig.~\ref{fig:ss_b}(b), which are
associated with MI behavior.

Even more significant is the behavior of the distribution of the average
site occupation number $P_{n}\left(n_{i}\right)$, shown in Fig.~\ref{fig:s1_todosb}(b).
As $\Delta$ increases, it is clearly seen that the distribution gradually
evolves into essentially isolated peaks centered around the integer
values (1 through 4 for $\Delta/U_{0}=16$). The presence of sites
with different integer occupations in a non-SF phase is the hallmark
of the BG, cf. Fig.~\ref{fig:p_n}(a).

\begin{figure}
\includegraphics[scale=0.25]{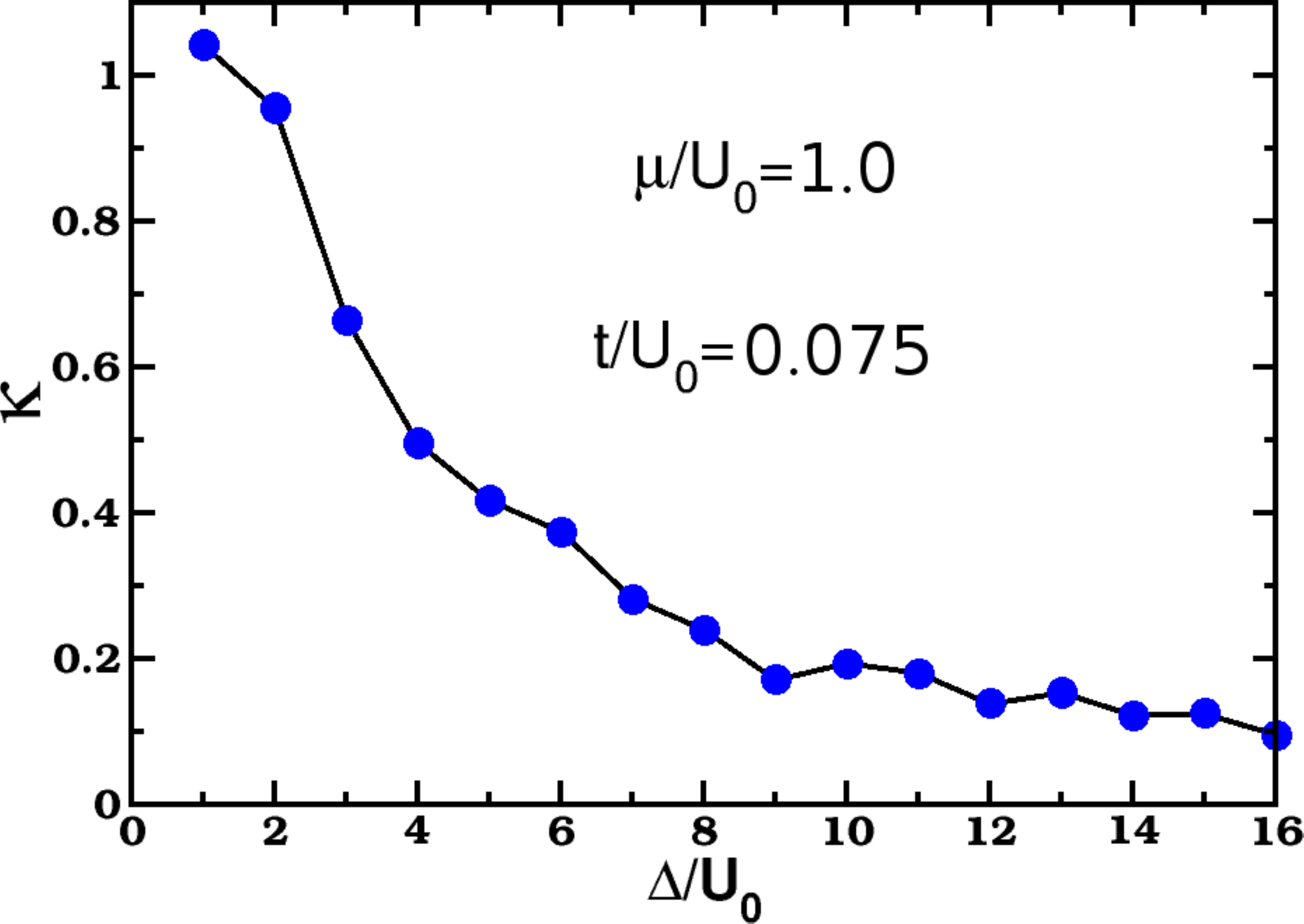}
\caption{\label{fig:compvsdisord}(Color online) 
The compressibility as a function of disorder strength. The chemical potential is fixed
at $\mu/U_{0}=1$ and the hopping amplitude at $t/U_{0}=0.075$.}
\end{figure}

In order to dissipate any doubt that the large disorder phase for this particular
choice of parameters is indeed a BG, we show in Fig.~\ref{fig:compvsdisord} the
compressibility as a function of disorder strength. Its value decreases with increasing
disorder but remains finite at the largest value analyzed ($\Delta/U_{0}=16$)
at which point the order parameter has already vanished, cf. Fig.~\ref{fig:s1_todosa}(a).

This shows that, though more robust against randomness, the even-numbered
MI lobes can also be wiped out and transformed into BG phases with
sufficiently large disorder. It follows that, for large enough values
of $\Delta$, only the SF and the BG phases survive, as had been previously
observed in the spin-zero case \cite{bissbort09,bissbort10}.

\section{Conclusions}

\label{sec:conclusion}

We have analyzed the stochastic mean field theory of the disordered
spin-1 Bose-Hubbard and discussed its physical properties as a function
of the hopping amplitude, the chemical potential and the disorder
strength. Although the model exhibits strong similarities with its
spin-zero counterpart, several differences stand out. There is a clear
difference in the behavior of the odd- and the even-numbered MI lobes.
The latter are much more robust with respect to the introduction of
disorder. As a result, there is a sizable portion of the parameter
space in which only even-numbered MI lobes exist, the odd-numbered
ones having been transformed into a BG. The BG insulator is characterized
by a finite compressibility and the presence of sites with different
occupations, like the spin-zero case. However, unlike the spin-zero
case, different occupations give rise to different spins. Therefore,
the spin-1 BG is an inhomogeneous mixture of spin-0 and spin-1 composites
within the SMFT. Very similar behavior was obtained within the Gutzwiller
approach of reference \onlinecite{sanpera11}. We should stress that reference 
\onlinecite{sanpera11} employs two different approaches in the study of 
the disordered system. In one approach, which the authors call a
`probabilistic mean-field theory', 
only an average order parameter is considered in the description. 
This approach is a much poorer description than the present SMFT,
since it incorporates no spatial fluctuations 
and, in particular, does not exhibit a BG phase. Alternatively, they also
show a direct lattice calculation of the site-decoupled mean-field theory. 
This does have spatial fluctuations and describes all 3 phases. 
It includes spatial correlations of the local order parameter 
which are absent in our SMFT treatment and should therefore 
be considered a superior approach. However, direct comparison 
shows that the phase diagram and some physical properties we obtain
are almost exactly the same as the lattice calculation of reference \onlinecite{sanpera11},
highlighting that the much simpler SMFT already incorporates 
the most important correlations of the more complete treatment.

The presence of local composites with different total spin values
raises the important question of the spin correlations within
the spin-1 BG phase. As discussed in \onlinecite{imambekov2003} for the
clean system, spin correlations outside the scope of either the `site-decoupled'
mean-field theory or the Gutzwiller approach give rise to non-trivial
nematic ground states. It would be of great interest to incorporate
these into a description of the disordered system in order to investigate
the interplay between disorder-induced number fluctuations and quantum
inter-site correlations. Besides the possibility of a spin nematic,
the introduction of randomness could also potentially give rise to
spin-glass order, a Bose-spin-glass or quantum Griffiths phases \cite{Lin2003a,Igl'oi2005,Miranda2005,Vojta2006}.

Another direction deserving of further scrutiny is the case of ferromagnetic
interactions, $U_{2}<0$. In this case, we expect the MI lobes to
be characterized by the bosons aligning to form a maximum spin composite.
In the presence of strong enough disorder, spins of different sizes
are expected to form, rendering the problem of the ground state spin
structure even richer.

\section*{Acknowledgments}

This work was supported by CNPq through grants 304311/2010-3 (EM)
and 140184/2007-4 (JHW) and by FAPESP through grant 07/57630-5 (EM).

\end{document}